\documentclass[twocolumn,showpacs,superscriptaddress,longbibliography]{revtex4-1}
\usepackage{graphicx}
\usepackage{array}
\usepackage{amsmath}
\usepackage{xcolor}
\usepackage[colorlinks=true, citecolor={blue!50!black}, urlcolor={blue!50!black}, linkcolor = {blue!50!black}]{hyperref}
\usepackage{amssymb}
\usepackage{bm}
\usepackage{color}
\usepackage{bookmark}
\usepackage{tabularx}
\usepackage{mathtools}
\usepackage{microtype}
\usepackage{hhline}

\newcommand{\ket}[1]{|#1\rangle}

\newcommand{\vt}[1]{\mathbf{#1}}
\newcommand{\abs}[1]{\left\vert#1\right\vert}

\DeclareMathOperator{\de}{d\!}
\DeclareMathOperator{\re}{Re}

\begin{document}

\title{Conductance of a proximitized nanowire in the Coulomb blockade regime}
\author{B. van Heck}
\affiliation{Department of Physics, Yale University, New Haven, CT 06520, USA}
\author{R. M. Lutchyn}
\affiliation{Station Q, Microsoft Research, Santa Barbara, California 93106-6105, USA}
\author{L. I. Glazman}
\affiliation{Department of Physics, Yale University, New Haven, CT 06520, USA}
\date{\today}
\begin{abstract}
We identify the leading processes of electron transport across finite-length segments of proximitized nanowires and build a quantitative theory of their two-terminal conductance. In the presence of spin-orbit interaction, a nanowire can be tuned across the topological transition point by an applied magnetic field. Due to a finite segment length, electron transport is controlled by the Coulomb blockade. Upon increasing of the field, the shape and magnitude of the Coulomb blockade peaks in the linear conductance are defined, respectively, by Andreev reflection, single-electron tunneling, and resonant tunneling through the Majorana modes emerging after the topological transition. Our theory provides the framework for the analysis of experiments with proximitized nanowires, such as reported in Ref.~\cite{albrecht2015}, and identifies the signatures of the topological transition in the two-terminal conductance.
\end{abstract}

\maketitle

\section{Introduction}

The possibility of realizing topological superconductivity~\cite{read2000,kitaev2001}, an exotic electronic phase hosting Majorana zero-energy modes, sparked a great amount of theoretical and experimental activity~\cite{nayak2008,alicea2012,beenakker2013,leijnse2012,stanescu2013}. Much of this excitement can be attributed to the prediction that defects in topological superconductors carry Majorana zero-energy modes and obey non-Abelian braiding statistics~\cite{nayak2008}. The latter, combined with the presence of an extensive ground-state degeneracy, opens the possibility for topological quantum computation~\cite{nayak2008, dassarma2015}.

Theory predicts that topological superconductivity can be realized when a conductor with strong spin-orbit interaction~\cite{fu2008,fu2009,sau2010,alicea2010,lutchyn2010,oreg2010,wimmer2010,potter2010,lutchyn2011,duckheim2011,chung2011,cook2011, stanescu2011, potter2011,potter2012}, or alternatively a chain of magnetic atoms \cite{choy2011,nadj2013,nakosai2013,pientka2013,braunecker2013,klinovaja2013,vazifeh2013,nadj2014,kim2014,brydon2015,li2014,heimes2015,zhang2016}, is coupled to a conventional superconductor. Following theoretical proposals~\cite{lutchyn2010, oreg2010}, some signatures of Majorana zero-energy states have been reported in semiconductor nanowires coupled to an $s$-wave superconductor~\cite{mourik2012, rokhinson2012, das2012, deng2012, finck2012, churchill2013}. Recent improvements of the quality of superconductor-semiconductor interface has been achieved by fabricating nanowires with a semiconducting core (InAs) and an epitaxial superconducting shell (Al)~\cite{krogstrup2015}. Thanks to the high quality of the proximity effect, these nanowires revealed a ``hard" superconducting gap close to that of Al \cite{chang2015, higginbotham2015, albrecht2015}, while in the earlier experiments \cite{mourik2012,das2012,deng2012,finck2012,churchill2013} zero-bias features (signatures of Majorana zero-energy modes) coexisted with a smooth subgap background. This development made it possible to study the interplay of proximity-induced superconductivity and charging effects in the Coulomb blockade regime \cite{higginbotham2015,albrecht2015} which allows one to probe the nature of ground-state degeneracy and investigate finite-size effects. A recent experiment of Albrecht et al.~\cite{albrecht2015} reported the detection of the ground-state degeneracy splitting. This is the first systematic measurement of the ground-state degeneracy associated with Majorana zero modes and is a milestone event which brings us one step closer to topological quantum computation.

Another reason for the excitement generated by the Copenhagen experiment~\cite{albrecht2015} is the possibility to use semiconducting nanowires as gate-tunable junctions and Josephson elements. Indeed, the nanowire-based Cooper-pair box is a highly-tunable device and has potential applications in superconducting electronics~\cite{delange2015,larsen2015}. The nanowire junctions can be tuned between weak and strong tunneling regimes with a few transverse channels which is to be contrasted with the conventional tunnel junctions in metallic islands, having a large number of weakly transparent channels.  Thus, recent experiments on proximitized nanowires~\cite{chang2015,higginbotham2015,albrecht2015} allow for an exploration of a richer phase diagram than the one accessible with the conventional superconducting islands, see, e.g.,  Refs. \cite{eiles1993,lafarge1993}.

\begin{figure}
\begin{center}
\includegraphics[width=\columnwidth]{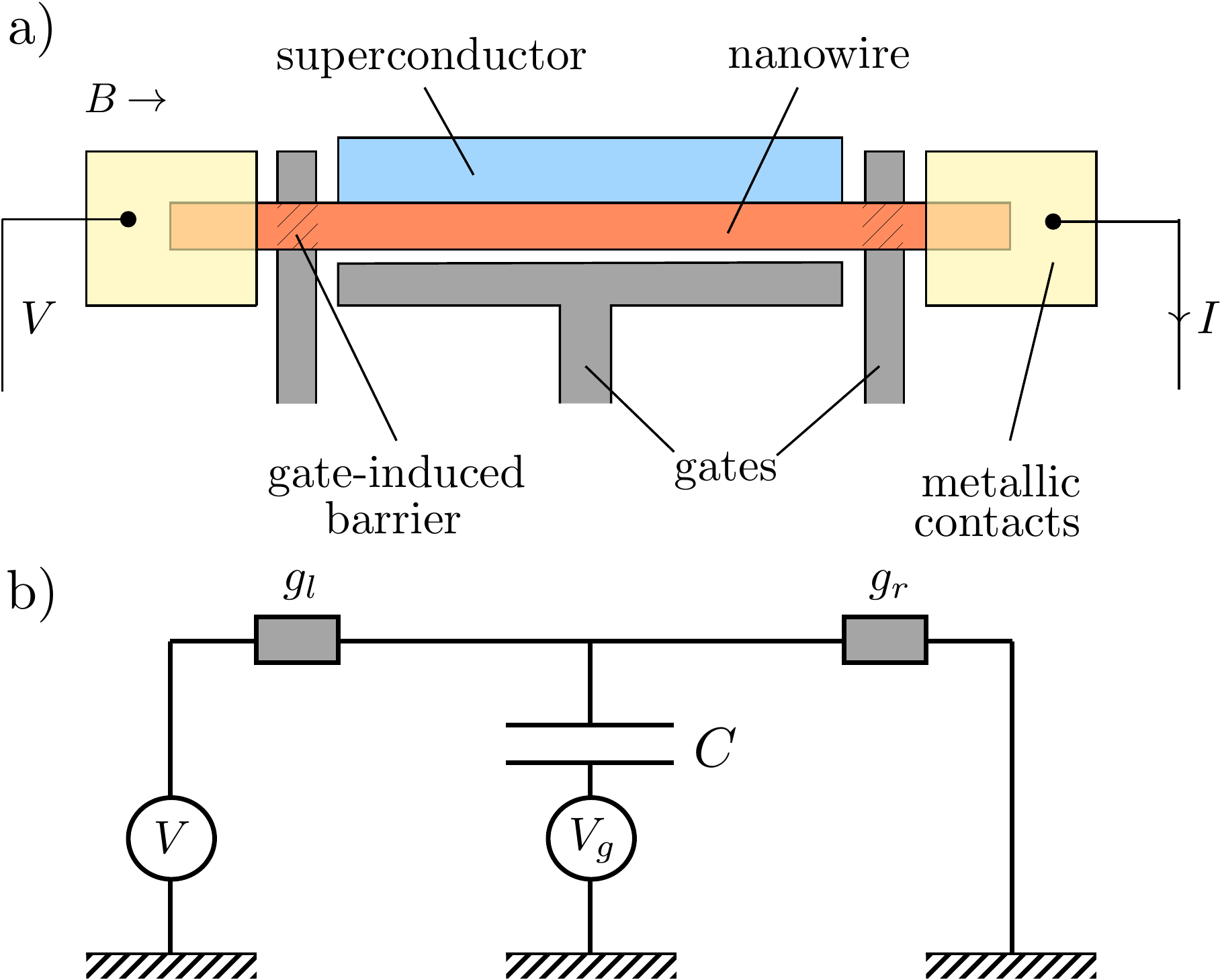}
\caption{\emph{Panel (a)}. Schematic drawing of the system under study. A semiconducting nanowire (e.g. InAs or InSb) is in proximity with a floating $s$-wave superconductor (e.g. Al). Underlying gates create tunnel barrier between the central part of the nanowire and two metallic contacts (e.g. Au), and control the electrostatic energy of the proximitized nanowire. A magnetic field $B$ can be applied parallel to the nanowire. \emph{Panel (b).} The electric circuit corresponding to panel (a). The nanowire and the superconductor form an almost isolated component of the circuit with a total capacitance $C$, and they are weakly connected to the leads via junctions of conductance $(2e^2/h) g_{l,r}$ with $g_{l,r}\ll 1$. The system is in the Coulomb blockade regime when $e^2/2C\gg eV, k_BT$. \label{fig:layout}}
\end{center}
\end{figure}

The Coulomb blockade of electron transport across a small conductor, see Fig.\ref{fig:layout} for a device layout, is associated with the electrostatic energy of electron charge the conductor carries.
The charging energy of a superconducting island discriminates between states with different number of electrons. That modifies the effect of BCS pairing on the excitations spectra, removing the gap for excitations if the electron number is odd. Further modifications of the ground and excited states come due to Majorana zero modes which inevitably appear in the case of $p$-wave pairing~\cite{kitaev2001}. The corresponding peculiarities in the spectra of fermionic systems were first considered in the context of nuclear physics~\cite{bohr1958,green1958}. The solid-state implementations pose a question as to how the same physics affects the electronic conduction across a superconducting island. The existing theories, which were addressing the $s$-wave pairing in islands of conventional superconductors~\cite{averin1992,hekking1993} and a basic model with $p$-wave pairing~\cite{fu2010} give qualitative, but not quantitative answers. This work fills the void, providing a quantitative theory applicable to proximitized nanowires connected to leads by single-channel junctions.

The type of the superconducting state in a wire is controlled by the competition between the effects of superconducting $s$-wave proximity and Zeeman splitting induced by an external magnetic field. The increase of the magnetic field results in the suppression of the induced by proximity $s$-wave gap; the gap eventually closes and re-opens in the $p$-wave channel. The evolution of the superconducting state prompts a sequence of the dominant electron transport mechanisms, from Andreev reflection, to single-electron tunneling, to resonant tunneling via Majorana states. We present a quantitative theory of the Coulomb blockade of the zero-bias conductance in each of these regimes, and discuss transitions between them, see Sections~\ref{sec:andreev}, \ref{sec:single_electron} and \ref{sec:majorana}. Sections~\ref{sec:model} and \ref{sec:rate_eqs} provide the formalism used to evaluate the conductance. The overview of the transport mechanisms and of our main results is given in Section~\ref{sec:Qual}.

\section{Qualitative discussion and main results}\label{sec:Qual}

\subsection{Dominant mechanisms of electron transport}

The electrostatic energy of the proximitized nanowire segment, see Fig.~(\ref{fig:layout}) varies with gate voltage as $E_c(N-n_g)^2$, where $N$ is the number of excess electrons, $E_c=e^2/2C$ is the charging energy, $C$ is the effective capacitance, and $n_g = C V_g/e$ is the gate voltage in dimensionless units.
At a small bias voltage and low temperature, $eV, k_BT\ll E_c$, the system is in the Coulomb blockade regime. The current is suppressed by the large charging energy, except at special values of the gate voltage $V_g$ where there is no energy cost associated with the transfer of charge through the system. The necessity for this resonant condition leads to the well-known Coulomb blockade oscillations: the occurrence, as a function of $V_g$, of high-conductance peaks separated by low-conductance valleys.

\begin{figure}
\begin{center}
\includegraphics[width=\columnwidth]{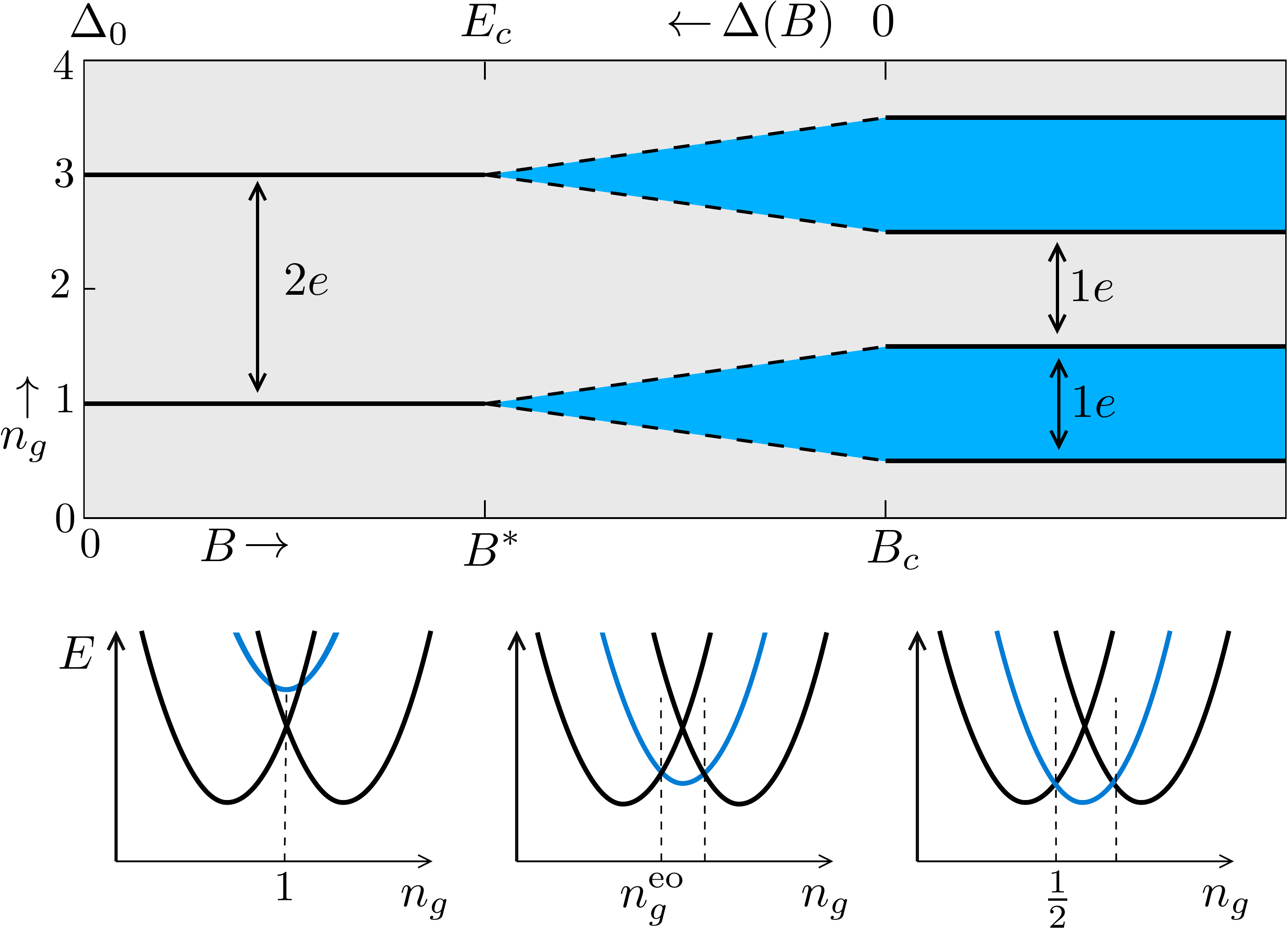}
\caption{Sketch of the position of the conductance peaks of Coulomb blockade oscillations as a function of magnetic field $B$, which controls the superconducting gap $\Delta(B)$, and dimensionless voltage $n_g = CV_g/e$, which controls the electrostatic energy $E_N = E_c(N-n_g)^2$. The ground state value of $N$ results from the competition of $\Delta$ and $E_N$ [see Eq.~\eqref{eq:gs_energy}], and peaks in conductance marks positions where the ground state value of $N$ changes. Only even values of $N$ are allowed for $B<B^*$ [$\Delta(B)>E_c$], while for $B>B^*$ [$\Delta(B)<E_c$] both even and odd values of $N$ are allowed. Odd ground state parity is marked by blue areas on the $(B, n_g)$ plane. The period of oscillations halves for $B>B_c$, either because the system has entered a topological superconducting phase with Majorana bound states, or because it has become metallic.\label{fig:sketch_peaks}}
\end{center}
\end{figure}

The Coulomb blockade oscillations reported in Ref.~\cite{albrecht2015} for the system in Fig.~\ref{fig:layout} exhibit the structure schematically shown in Fig.~\ref{fig:sketch_peaks}. There are three distinct types of behavior observed as the magnetic field is increased from zero. At weak field, the oscillations are periodic in gate voltage with period $2e/C$, corresponding to the charge of a Cooper pair. There is a single conductance peak within each period, achieved at odd integer values of the dimensionless voltage $n_g$. At a certain value $B^*$ of the magnetic field, within each period the conductance peak splits in two. At $B>B^*$, the positions of the two peaks within each period move away from each other with a shift approximately linear in field. At a second value of the magnetic field $B_c$, the fundamental voltage period of the oscillations becomes $e/C$. For $B>B_c$ the conductance peaks occur at half-integer values of $n_g$, independently of the magnetic field.

Qualitatively, this behavior can be interpreted in terms of the interplay between the superconducting gap $\Delta(B)$ and the charging energy $E_c$, as in the early experiments on the even-odd effect in superconducting islands \cite{eiles1993,lafarge1993}. Due to the superconducting pairing, the total ground state energy of the proximitized nanowire with $N$ excess electron charges depends dramatically on the parity of $N$:
\begin{equation}\label{eq:gs_energy}
E_\textrm{gs}(N) = E_c(N-n_g)^2 +
\begin{cases}
\Delta(B) & \textrm{if $N$ is odd,} \\ 0 & \textrm{if $N$ is even.}
\end{cases}
\end{equation}
For a given value of $n_g=CV_g/e$, the ground state is determined by minimizing $E_\textrm{gs}(N)$ as a function of $N$. Let us assume, now and in the rest of the paper, that at zero magnetic field $\Delta(0)>E_c$. Then, at $B=0$, the ground state always corresponds to an even value of $N$. Ground state degeneracy points at which $E_\textrm{gs}(N) = E_\textrm{gs}(N+2)$ occur when $n_g$ is an odd integer, explaining the $2e$-periodicity of the Coulomb blockade peaks. At these charge degeneracy points, the leading conduction mechanism is the resonant transfer of electron pairs through the nanowire, mediated by Andreev reflection processes at the tunnel junctions.

Upon increasing $B$, the gap $\Delta(B)$ starts to decrease until one encounters the value $B^*$ for which $\Delta(B^*)=E_c$. For $B>B^*$, the gap is smaller than the charging energy and therefore the ground state parity may change. For even values of $N$, the transition to an odd state takes place when $E_\textrm{gs}(N)=E_\textrm{gs}(N+1)$, which happens if $n_g = N+n_g^\textrm{eo}$ with
\begin{equation}\label{eq:ngeo}
n_g^\textrm{eo} = \frac{\Delta(B)+E_c}{2E_c}\,.
\end{equation}
The odd state remains the ground state until a second degeneracy point $E_\textrm{gs}(N+1)=E_\textrm{gs}(N+2)$ is encountered, which happens at $n_g = N+2-n_g^\textrm{eo}$. This explains the occurrence of two Coulomb peaks in the conductance within the same $2e$ interval. At these degeneracy points, the leading conduction mechanism is the resonant transfer of single electrons rather then electron pairs \cite{averin1992}. The odd Coulomb valleys extend over a voltage range proportional to $1-\Delta(B)/E_c$, which grows with the increasing value of $B-B^*>0$. Eventually, at $B=B_c$ the odd and even valleys become of the same length (see Fig.~\ref{fig:sketch_peaks}).

At field $B=B_c$, the superconducting gap closes, $\Delta(B_c)=0$. The halving of the period of the zero-bias conductance oscillations with $n_g$ at $B>B_c$ indicates the absence of an even-odd effect in the ground state of the system. This is the expected behavior of both a metallic island in the normal state, and a topological superconductor, where a single fermionic quasiparticle can occupy a zero-energy state ``shared'' by the two Majorana bound states. In the first interpretation, $B_c$ must be the critical magnetic field which destroys superconductivity, while in the second $B_c$ is identified with the critical field of the topological transition. The theory~\cite{lutchyn2010,oreg2010} predicts that in proximitized nanowires with strong spin-orbit coupling, the increase of an external magnetic field $B$ can induce a topological phase transition signaled by the closing of the proximity-induced superconducting gap $\Delta(B)$ at a critical value $B=B_c$. Gap reopens at higher field, accompanied by the appearance of Majorana zero-energy bound states at the ends of the proximitized nanowire. These states facilitate resonant electron tunneling (dubbed ``teleportation'' in Ref.~\cite{fu2010}) at the charge degeneracy points.

In a finite-length wire, the Majorana modes localized at the opposite ends hybridize, resulting in an exponentially small  splitting of the ground-state degeneracy $\delta E \propto \exp(-L/\xi)$~\cite{kitaev2001,cheng2009}. Here $L$ and $\xi$ are the nanowire length and the effective superconducting coherence length, respectively. This splitting results in the corresponding small shift of the Coulomb blockade peak positions. The shift and its dependence on $L$ are perceived as one of the manifestations of the topological phase~\cite{albrecht2015}. The prominence of the shift depends on how sharp the Coulomb blockade peaks in conductance are. Finding the magnitudes and shapes of the peaks for each of the described mechanisms of conduction is the goal of our quantitative theory.

\subsection{Relevant energy scales and simplifications}

We begin by discussing the typical energy scales and relations between them for a conventional-semiconductor wire segment proximitized by a conventional superconductor, such as InAs/Al system experimented with in Ref.~\cite{albrecht2015}. We use units with $\hbar=k_B=1$.

The gap induced in the nanowire, $\Delta(B)$, depends on the applied magnetic field. This dependence is controlled by the competition between the proximity-induced $s$-wave superconductivity and the Zeeman effect (one may neglect the orbital effect of the magnetic field due to the small diameter of the wire). At some critical value, $B=B_c$, the gap closes. Thanks to the presence of spin-orbit coupling, and provided the Fermi level in the nanowire lies within the Zeeman gap, the gap re-opens at fields $B>B_c$, at which the system is in the topological phase with effective $p$-wave pairing. For brevity, we will make the $\Delta (B)$ dependence implicit by skipping the argument $B$ and will write $\Delta(B)\equiv \Delta$, unless required by clarity. Furthermore, we will denote $\Delta(0)\equiv\Delta_0$.

The induced gap value is naturally limited by the gap $\Delta_\textrm{Al}$ in the source of the proximity, $\Delta_0<\Delta_\textrm{Al}$. If the source is a narrow superconducting shell around the wire, then one may neglect the suppression of $\Delta_\textrm{Al}$ due to the orbital effect of a magnetic field applied along the wire. Having in mind applications related to Majorana physics, we will assume the $g$-factor of the semiconducting nanowire, $g$, to strongly exceed the $g$-factor in the superconductor. That allows us to disregard the Zeeman effect in the superconductor and neglect the $B$-dependence of $\Delta_\textrm{Al}$ in a range of field containing the interesting value $B=B_c$.

The above assumptions are adequate for a range of materials and geometries. For example, in the experiment~\cite{albrecht2015} with the InAs/Al nanowire system, the superconducting gap in the aluminium shell was $\Delta_\textrm{Al}\approx 180\,\mu$eV, and it persisted until $\approx 1$ T. We estimate the zero-field induced gap value to be $\Delta_0\approx 110-150\,\mu$eV. The induced gap closed and re-opened at $B\approx0.1-0.2$ T.

The charging energy $E_c$ may vary depending on the length of the nanowire segment between the two tunnel barriers, as well as other details regarding the layout of metallic gates surrounding the nanowire. Without loss of generality, we will focus on values of $E_c$ such that $E_c<\Delta_0$, which would allow one to explore all the three regimes of Fig.~\ref{fig:sketch_peaks} upon varying the magnetic field $B$, as observed experimentally~\cite{albrecht2015}. At $T\approx 50$ mK (that is, $T\approx 5\,\mu$eV), this constraint is fully compatible with the Coulomb blockade regime which requires $T\ll E_c$.

An important energy scale in determining the transport properties of the system is the level spacing $\delta$ for states in the proximitized nanowire around the Fermi level. The latter is larger than the level spacing $\delta_\textrm{Al}$ in the Al shell, $\delta > \delta_\textrm{Al}$, for two reasons. First, the Fermi wavelength in the superconductor is much smaller than that of the semiconductor, thus leading to a much larger number of transverse channels in the Al shell. Second, the Fermi velocity in the semiconductor is much lower than that of the superconductor.

We may estimate the two level spacings as follows. For $\delta_\textrm{Al}$, using the known value of the density of states in Al \cite{higginbotham2015} and a shell volume of $\sim 10^5$ nm$^3$ (corresponding to a rather thin shell of 1 $\mu$m$\times$10 nm$\times$10 nm), one may estimate $\delta_\textrm{Al}\lesssim 1\,\mu$eV. Regarding the nanowire, the goal of engineering Majorana  states calls for a high value of the spin-orbit interaction $\alpha$. While there is no certain knowledge regarding the value of $\alpha$ in the InAs/Al nanowires, a conservative estimate \cite{albrecht2015} is $\alpha \approx\!10^4$ m$/$s $\approx 10\,\mu$eV$\cdot\mu$m. This value gives the scale for the Fermi velocity in the limit of a low Fermi energy in the nanowire, favorable for the opening of a Zeeman gap at the Fermi level, and thus it also sets the relevant scale $\pi \alpha /L$ for the level spacing of a segment of length $L$ of a ballistic, single channel nanowire. The requirement for a homogeneous wire sets a limit on $L$, and, although there are no fundamental limitations, the current typical length is $L\approx 1 \mu$m, for which we obtain $\pi \alpha / L \gtrsim\,30\,\mu$eV. Note, however, that this estimate should be considered as an upper bound on $\delta$, because in the presence of the superconducting shell the level spacing is renormalized, and in fact can be considerably reduced due to the strong hybridization with states in Al (see Appendix~\ref{app:level_spacing}).

In fact, we will assume that $\delta \ll \Delta$, which is the favorable condition for a clear detection of Majorana bound states. Indeed, a large value of the level spacing would also imply in the topological phase at $B>B_c$ a significant finite-size coupling between Majorana bound states at the opposite ends of the wire. The same condition also guarantees that the induced level spacing of states right above the gap is small, $\delta^2/\Delta \ll T$, which, in turn, guarantees that for $B<B_c$ the transport properties are determined by more than a few states in the nanowire, so that we do not need to worry about the fine details of such states. The two conditions $\delta\ll \Delta$ and $\delta^2/\Delta\ll T$ are assumed to be valid at any value of the magnetic field, provided one is not too close to the phase transition when $\Delta(B)$ vanishes. Hence, both $\delta/\Delta$ and $\delta/\sqrt{\Delta T}$ are small parameters for the theory and the accuracy of our calculations is increasing with the length of the nanowires.

The level spacing $\delta$ and the energy gap $\Delta$ also set the characteristic energy scale for quasiparticle poisoning in a superconductor, the \emph{poisoning temperature} \cite{matveev1994},
\begin{equation}\label{eq:Tp}
T_p = \frac{\Delta}{\ln (\sqrt{2\pi}\Delta/\delta)}\,.
\end{equation}
In an isolated superconductor without charging energy, the number of thermal quasiparticles is large if $T\gg T_p$, and negligibly small if $T\ll T_p$. Note that $T_p<\Delta$, due to the large spectral weight of odd parity states - which is proportional to the number of single particle states and hence to the volume of the system - with respect to the even parity states. In the presence of charging energy, the condition $T\lesssim T_p$ (or, even better, $T\ll T_p$) still ensures that the number of quasiparticles in the superconducting dot is of order one. As an example, plugging $\Delta_0\approx 130\,\mu$eV and $\delta\approx 5\,\mu$eV in Eq.~\eqref{eq:Tp}, one obtains $T_p\approx 30\,\mu$eV. Note that the smaller $\delta$, the more stringent the condition $T\ll T_p$ is on the temperature $T$.

Let us now discuss the properties of the contacts. In the relevant scenario, the proximitized nanowire is contacted by tunnel junctions with a small conductance $G_{l,r} \ll 2 e^2 / h$. In a semiconducting junction this level of conductance is achieved by one or few conducting channels, as opposed to the hundreds of low-transmission channels of a metallic junction. Aiming at setups optimized for the observation of the Majorana bound states, we model the junctions as single-channel point contacts. Their strength is measured by the dimensionless conductances $g_{l,r} = (h/2e^2) G_{l,r}$ which, we assume here, are much smaller than one. In the presence of the contacts, quantum states in the nanowire acquire a finite lifetime, with the typical broadening of a single-particle state with energy close to the Fermi level given by $(g_l + g_r)\delta$.

The presence of the small parameters $g_l, g_r, \delta/\Delta$ and $\delta/\sqrt{\Delta T}$ allows us to treat the problem within a simple perturbative approach, with the aim to compute the conductance perturbatively to the leading non-zero order in $g_l, g_r$ and $\delta$. The perturbative approach breaks down at very low temperatures, where many-body effects related to quantum fluctuations of electric charge become relevant \cite{houzet2005}. At low conductances, the energy scale governing the onset of such many-body effects is the ``charge-Kondo'' temperature $T_K = E_c\,\exp[-\tfrac{\pi^2}{2(g_l+g_r)}]$ \cite{garate2011}. In this paper, we neglect many-body effects, working in the limit $T_K/T\to 0$.

\subsection{Overview of main results}

Before going into the details of the calculations, we summarize here the main results of our theory regarding the Coulomb peaks in the different regimes of Fig.~\ref{fig:sketch_peaks}.

The case $B=0$ and $\Delta_0>E_c$ is discussed in Sec.~\ref{sec:andreev}, where we reproduce known results on the Coulomb blockade of Andreev reflection \cite{hekking1993}. The system exhibits $2e$-periodic conductance peaks with amplitude [see Eq.~\eqref{eq:G_2e}]
\begin{equation}\label{eq:g_peak_2e}
G^\textrm{peak}_\textrm{2e}\sim \frac{2e^2}{h}\,\frac{g^2_lg^2_r}{g^2_l+g^2_r}\,.
\end{equation}
The conductance peak is symmetric around the peak position $n_g=1$, with activated tails and width proportional to $T/E_c$.

In Sec.~\ref{sec:single_electron} we compute the differential conductance in the situation $\Delta(B)<E_c$, where single-electron tunneling is the relevant transport process. In this intermediate regime, the conductance has two rather dim peaks in a $2e$-interval, with [see Eq.~\eqref{eq:sequential_peak}]
\begin{equation}\label{eq:g_peak_e}
G^\textrm{peak}_\textrm{e} \sim \frac{e^2}{h}\,\frac{g_lg_r}{g_l+g_r}\,\frac{\delta}{T}\,.
\end{equation}
The peak vanishes in the limit $\delta\to 0$ (long wire). The peak position is shifted from the expected value $n_g^\textrm{eo}$ of Eq.~\eqref{eq:ngeo} towards smaller values, thus enlarging the region in Fig.~\ref{fig:sketch_peaks} with odd ground state parity, and the peak width is proportional to $(\Delta/E_c)\,(T/T_p)$. As we show in Sec.~\ref{sec:elastic}, the peak may develop a marked asymmetry due to elastic co-tunneling processes \cite{averin1992}, which yield larger conductance on the even side of the transition. This contribution to the conductance has a weak temperature dependence, and its visibility with respect to the sequential tunneling contribution is enhanced at low temperatures and large level spacings, consistent with experimental observations \cite{higginbotham2015}.

The single-electron tunneling peak is sensitive to the level spacing, unlike the Andreev peak, but it is of lower order in the conductances $g_l, g_r$. We may compare the heights of the Andreev and single-electron peaks by taking the ratios of Eq.~\eqref{eq:g_peak_e} and Eq.~\eqref{eq:g_peak_2e}. In the limit of symmetric point contacts, $g_l=g_r\equiv g$, we obtain $G^\textrm{peak}_\textrm{e}/G^\textrm{peak}_\textrm{2e}\sim (\delta/T)(1/g)$.  This estimate is valid for $\Delta_0\gg E_c$; a more accurate estimate can be taken by using Eq.~\eqref{eq:G_2e} for the Andreev peak, which includes the detailed dependence of $G_\textrm{2e}$ on $\Delta_0$ and $E_c$.

In Sec.~\ref{sec:majorana} we consider the case $B>B_c$, where the conductance oscillations are $1e$-periodic due to resonant tunneling via Majorana bound states \cite{fu2010}. The conductance achieves a maximum value [see Eqs.~\eqref{eq:g_maj} and \eqref{eq:g_maj_finite_T}]
\begin{equation}\label{eq:g_peak_maj}
G_\textrm{Maj}^\textrm{peak}\sim \frac{e^2}{h}\,\frac{g_l g_r}{g_l+g_r}\,\frac{\Delta}{8T}.
\end{equation}
The above equation is valid for temperatures $T\gg \Gamma_\textrm{Maj}$ where $\Gamma_\textrm{Maj}=(g_l+g_r)\,\Delta/8\pi$ is the broadening of the zero-energy Majorana state, see Eq.~\eqref{eq:Gamma_Maj}. In the opposite limit $T\ll\Gamma_\textrm{Maj}$, the Majorana peak height becomes independent of $T$ and equal to $(4e^2/h)\,g_lg_r/(g_l+g_r)^2$, which yields the conductance quantum $e^2/h$ when $g_l=g_r$. The width of the peak is proportional to $\max(T, \Gamma_\textrm{Maj})/E_c$. Finally, it is interesting to compare the peak height in the Majorana regime to that in single-electron tunneling regime. Taking the ratios of Eq.~\eqref{eq:g_peak_e} and Eq.~\eqref{eq:g_peak_maj}, we obtain $G^\textrm{peak}_\textrm{e}/G^\textrm{peak}_\textrm{Maj}\sim \delta/\Delta$, independent of the conductances of the point contacts.

\section{Model Hamiltonian}\label{sec:model}

In the weak-tunneling limit, one may adopt the tunneling Hamiltonian formalism. The total Hamiltonian for the system reads as
\begin{equation}\label{eq:h_tot}
H_\textrm{tot} = H_\textrm{leads}+H_\textrm{wire}+H_\textrm{tunn}\,.
\end{equation}
The first term is the Hamiltonian for the leads,
\begin{equation}\label{eq:h_leads}
H_\textrm{leads}=\sum_{j,p\sigma} \xi_p\,c^\dagger_{j,p\sigma}\,c_{j,p\sigma}
\end{equation}
Here, $j=l,r$ labels the two leads, $p$ labels different single-particle orbitals with energy $\xi_p$, and $\sigma$ is the spin label. We assume the g-factor of leads to be small and dispense with the spin polarization in the leads. This is not restrictive, as long as the spin polarization of leads remains small \footnote{Note that if there is a long segment of semiconducting nanowire between the barriers and the leads, then a finite magnetic field may gap out half of the incoming and outgoing modes at the Fermi level, thus making the leads effectively spinless. This circumstance would affect some of the results quantitatively, in particular with regard to the magnetic-field dependence of the Andreev reflection amplitude computed in Sec.~\ref{sec:andreev} [see discussion below Eq.~\eqref{eq:G_2e}].}.

The Hamiltonian for the proximitized nanowire is
\begin{equation}
H_\textrm{wire}=E_c\,(\hat{N}-n_g)^2 + H_\textrm{BCS} \,.\label{eq:h_wire}
\end{equation}
The first term is the electrostatic energy [cf. Eq.~\eqref{eq:gs_energy}], with $\hat{N}$ being the electron number operator for the proximitized nanowire. The second term is the microscopic BCS Hamiltonian for the proximitized nanowire. By solving a system of Bogoliubov-de Gennes equations \cite{degennes}, $H_\textrm{BCS}$ can always be brought to a diagonal form
\begin{equation}\label{eq:h_bcs}
H_\textrm{BCS}=\sum_\alpha \epsilon_\alpha \gamma^\dagger_\alpha \gamma_\alpha + \textrm{const}
\end{equation}
with Bogoliubov quasiparticle operators $\gamma_\alpha$ obeying conventional fermionic commutation relations, where the label $\alpha$ may include different quantum numbers such as orbital and spin indices; $\epsilon_\alpha$ is the energy of a quasiparticle in the state labeled by $\alpha$. In terms of the quasiparticle operators, the electron field operator $\Psi$ at a position $\vt{r}$ and spin $\sigma$ is
\begin{equation}\label{eq:Psi}
\Psi(\vt{r},\sigma) = \sum_\alpha u_\alpha(\vt{r},\sigma) \gamma_\alpha + v^*_\alpha(\vt{r},\sigma)\,\gamma^\dagger_\alpha\,,
\end{equation}
where $u_\alpha$ and $v_\alpha$ are the solutions of the Bogoliubov-de Gennes equations.

While there is no fundamental obstacle in determining the energies $\epsilon_\alpha$ and the eigenfunctions $u^\alpha, v^\alpha$, this would require a complete microscopic description of the nanowire which would necessarily include several competing physical effects. The magnetic field affects the single particle states via Zeeman and orbital effects, and due to the strong spin-orbit coupling the induced pairing will be a mixture of singlet and triplet components. Furthermore, one may have to include the effects due to disorder, interface scattering and confinement potentials. Clearly we do not aim to achieve a comprehensive analysis of these effects, many of them already thoroughly investigated in the extensive literature on Majorana nanowires \cite{stanescu2011,brouwer2011,lutchyn2012,prada2012,rainis2013,sau2013,takei2013,stanescu2013,nijholt2015,vuik2016}. Rather, we find that it is possible to identify the leading transport mechanism and compute the parametric dependence of the conductance on $\Delta(B)$ and $n_g$ without such a fine level of details. Therefore, in the following sections we will only consider appropriate, simple limits of Eq.~\eqref{eq:h_wire}.

Due to the hybrid nature of the proximitized nanowire, and to the large mismatch between the Fermi wavelengths in the semiconductor and in the superconductor, there are two qualitatively different regions in the energy spectrum of Eq.~\eqref{eq:h_bcs}. The wave functions of states with energy $\epsilon_\alpha<\Delta_\textrm{Al}$ are effectively one-dimensional, being mostly localized in the semiconducting nanowire, while those of states with energy $\epsilon_\alpha>\Delta_\textrm{Al}$ are, on the other hand, three-dimensional and mainly localized in the superconductor. This second region of the spectrum is characterized by the much smaller level spacing $\delta_\textrm{Al}$.

The last term in Eq.~\eqref{eq:h_tot} is the tunneling Hamiltonian for the two point contacts. In view of the above considerations, it can be written as
\begin{align}\label{eq:h_tunn}\nonumber
\!H_\textrm{tunn} \!&=\sum_{j,p\sigma\alpha}W_j\left[\theta(\Delta_\textrm{Al}-\epsilon_\alpha)+\sqrt{\nu/\nu_\textrm{Al}}\,\theta(\epsilon_\alpha-\Delta_\textrm{Al})\right]\\
&\times\left[u^*_\alpha(\vt{r}_j,\!\sigma)\gamma^\dagger_\alpha \!+\! v_\alpha(\vt{r}_j,\!\sigma)\gamma_\alpha\right]\phi_p(\vt{r}_j)\,c_{p\sigma} \!+\! \textrm{H.c.}
\end{align}
Here, $\nu$ is the 1D density of states in the semiconducting nanowire, $\nu_\textrm{Al}$ is the 3D density of states in the superconductor, $\theta(\,\cdot\,)$ is the Heaviside step function, and finally $u_\alpha(\vt{r}_j,\sigma)$,  $v_\alpha(\vt{r}_j,\sigma)$ and $\phi_p(\vt{r}_j)$ are the normalized wave functions of the eigenstates of $H_\textrm{BCS}$ and $H_\textrm{leads}$ respectively, evaluated at the positions $\vt{r}_j$ of the contacts. Finally, the tunneling matrix elements quantities $W_j$ can be related to the conductances $g_j$ of the point contacts via the equation
\begin{equation}\label{eq:Wj}
W_j = \frac{1}{2\pi}\,\sqrt{\frac{g_j}{\nu_j\nu}}\,
\end{equation}
where $\nu_j$ is the 3D density of states in the leads. The three density of states $\nu, \nu_\textrm{Al}$ and $\nu_j$ are evaluated at the Fermi energy and without accounting for spin degeneracies. In particular, for a single-channel nanowire with low Fermi energy, $\nu\approx 1/\pi \alpha$, with $\alpha$ the spin-orbit interaction strength. In writing Eq.~\eqref{eq:Wj}, we have assumed that the conductances $g_j$ are not influenced by the magnetic field $B$ or by the presence of the superconducting shell.

In Eq.~\eqref{eq:h_tunn}, we have further assumed that the amplitude for tunneling through the contacts is spin independent, and we have used the fact that the spatial wave function $\phi_p(\vt{r}_j)$ for the states in the leads is the same for both spin directions; it satisfies $|\phi_p(\vt{r}_j)|^2 = 1/\Omega_j$, where $\Omega_j$ is the volume of lead $j$; in some intermediate formulae, we will make use of the lead level spacings $\delta_j = (\nu_j\Omega_j)^{-1}$, which we assume to be infinitesimally small and which drop out of final results.

The normalization factor of the wave functions $u_\alpha(\vt{r}_j,\sigma)$ [and, equivalently, $v_\alpha(\vt{r}_j,\sigma)$] differs for the two regions of the spectrum. For states with energy $\epsilon_\alpha<\Delta_{\rm Al}$, $|u(\vt{r}_j,\sigma)|^2\sim Z(\epsilon_\alpha)/L$, where $Z(\epsilon_\alpha)$ is a factor which accounts for the reduced weight of low-energy single-particle states in the nanowire due to the coupling to the superconductor [$0<Z(\epsilon_\alpha)<1$], see Appendix~\ref{app:level_spacing}. On the other hand, for states with $\epsilon_\alpha>\Delta_{\rm Al}$, $|u(\vt{r}_j,\sigma)|^2\sim 1/\Omega_\textrm{Al}$. Note that the precise value of $u_\alpha(\vt{r}_j,\sigma)$, $v_\alpha(\vt{r}_j,\sigma)$ is subject to mesoscopic fluctuations due to disorder, to the microscopic details of the junction, or both \cite{jalabert1992,prigodin1993,aleiner1996,beenakker1997,alhassid2000,aleiner2002}. In the context of the this work, mesoscopic fluctuations of the conductance turn out to be unimportant, except for the elastic co-tunneling calculation in Sec.~\ref{sec:elastic} and for the resonant tunneling through the Majorana bound states of Sec.~\ref{sec:majorana}, in which case we address the ensemble-averaged quantities.

\section{Rate equations}\label{sec:rate_eqs}

In order to compute the conductance, it is first convenient to project the wire Hamiltonian~\eqref{eq:h_wire} on a manageable subset of the entire Fock space on which it acts. The periodicity of the Coulomb blockade oscillations allows us to restrict the dimensionless gate voltage to an interval $n_g\,\in\,[N,N+2]$. In this voltage range, and given that $T\ll E_c$, we may restrict the analysis to the eigenstates of the operator $\hat{N}$ in Eq.~\eqref{eq:h_wire} with eigenvalues $N, N+1, N+2$. Moreover, the condition $T\lesssim T_p$ allows us to neglect all states with more than one excited quasiparticle in the nanowire. This leaves us with two states with even parity differing by one Cooper pair, which we denote $\ket{0}$ and $\ket{2}$, and a (large) set of odd-parity states which we denote $\ket{1; \alpha}$.

Let $P_0$, $P_\alpha$ and $P_2$ be the probability for the system to be in each of these states. In the presence of a finite bias voltage $V$ between the two contacts, the occupation probabilities for states in the wire can change in time due to the transfer of electrons between the wire and the leads. Close to degeneracy points, the charge transfer is dominated by incoherent processes and, as usual for Coulomb blockade systems, we can describe the time evolution of $P_0$, $P_\alpha$ and $P_2$ in terms of a system of rate equations \cite{glazman1988,beenakker1991,averin1991}. Each transition from an initial state $\ket{i}$ to a final state $\ket{f}$ of the wire is characterized by a transition rate $\Gamma_{i\to f}$, obtained using Fermi's Golden Rule. The amplitude for the process $\ket{i}\to\ket{f}$ can be computed perturbatively in the tunneling Hamiltonian, Eq.~\eqref{eq:h_tunn}, projected on the low-energy Hilbert space spanned by the states $\ket{0}, \ket{1;\alpha}, \ket{2}$.

The appropriate system of rate equations can be written by requiring, for
each state $\ket{i}$, a balance between transition from $\ket{i}$ and transition to $\ket{i}$. In our case, the resulting system of rate equations reads as
\begin{subequations}\label{eq:rate_equations}
\begin{align}\nonumber
\dot{P}_0 &= - \left(\Gamma_{0\to 2} + \sum_\alpha \Gamma_{0\to
\alpha}\right)P_0 + \Gamma_{2 \to 0} \,P_{2}\, + \\
&\qquad + \sum_\alpha \Gamma_{\alpha \to 0} P_\alpha\,,\\\nonumber
\dot{P}_{2} & =- \left(\Gamma_{2\to 0} + \sum_\alpha\,\Gamma_{2\to
\alpha}\right)P_{2} +\Gamma_{0\to 2}\, P_0\,+ \\
&\qquad + \sum_\alpha\Gamma_{\alpha\to 2}P_\alpha\,,
\\\nonumber
\dot{P}_\alpha & = - \left(\Gamma_{\alpha\to 0} + \Gamma_{\alpha\to
2}\right)P_\alpha + \Gamma_{0\to \alpha}\,P_0 + \\
&\qquad + \Gamma_{2\to \alpha} P_{2}\,.
\end{align}
\end{subequations}
The transition rates appearing in Eq.~\eqref{eq:rate_equations} may be divided in two types. $\Gamma_{0\to \alpha}, \Gamma_{\alpha\to0}, \Gamma_{2\to \alpha}$, and $\Gamma_{\alpha \to 2}$ all correspond to transitions which change the number of electrons in the wire by one. On the other hand, $\Gamma_{0\to 2}$ and $\Gamma_{2\to 0}$ correspond to Andreev reflection processes, which change the number of electrons in the wire by two either by removing or adding a pair. We postpone the detailed calculation of the different transition rates to
the next Sections. For the moment, we just note that each transition rate appearing in Eq.~\eqref{eq:rate_equations} is the sum of two contributions from the left and right contacts,
\begin{equation}
\Gamma_{i\to f} = \Gamma^l_{i\to f}+\Gamma^r_{i \to f} = \sum_j \Gamma^j_{i \to
f}\,.
\end{equation}
We are interested in computing the current in the steady-state achieved in the presence of the dc bias voltage $V$. The steady-state occupation probabilities can be determined by solving the linear system of equations obtained from
Eq.~\eqref{eq:rate_equations} by setting $\dot{P}_0=0$, $\dot{P}_\alpha=0$, and $\dot{P}_2=0$, together with the normalization condition
\begin{equation}\label{eq:normalization_probabilities}
P_0+ \sum_\alpha P_\alpha+ P_2 = 1\,.
\end{equation}
The linear system thus obtained has a unique solution, which can be presented in the following form:
\begin{subequations}\label{eq:steady_state_full}
\begin{align}
P_0 &= \frac{1}{B}\,\,,\\
P_2 &= \frac{A}{B}\,\,,\\
P_\alpha & = \frac{1}{B}\,\frac{\Gamma_{0\to \alpha} + A\,\Gamma_{2\to
\alpha}}{\Gamma_{k\to 0} + \Gamma_{\alpha\to 2}}\,,
\end{align}
with
\begin{align}
A &= \left(\Gamma_{0\to 2} - \sum_\alpha\frac{\Gamma_{0\to \alpha}\Gamma_{\alpha \to 2}}{\Gamma_{\alpha \to 0}+\Gamma_{\alpha \to 2}}\right)\\
&\times \left(\Gamma_{2\to 0} + \sum_\alpha \Gamma_{2\to \alpha} + \sum_\alpha \frac{\Gamma_{\alpha\to
2}\Gamma_{2\to \alpha}}{\Gamma_{\alpha\to 0}+\Gamma_{\alpha\to 2}}\right)^{-1}\,,\nonumber
\end{align}
and
\begin{align}\nonumber
B &= 1 + A\,\left(1+\sum_\alpha\,\frac{\Gamma_{2\to \alpha}}{\Gamma_{\alpha\to
0}+\Gamma_{\alpha \to 2}}\right) + \\ & \qquad + \sum_\alpha\,\frac{\Gamma_{0\to
\alpha}}{\Gamma_{\alpha\to 0}+\Gamma_{\alpha\to 2}}\,.
\end{align}
\end{subequations}
Although the full solution appears rather complicated, we will see that
depending on the values of $\Delta, E_c$ and $n_g$ one may often neglect some of the transition rates due to energy considerations. Hence, we will be mainly
concerned with some simple limits of the solution.

Once the transition rates and occupation probabilities for the different states of the wire are known, the current in the steady state can be easily computed. We write the total current as the sum of two contributions,
\begin{equation}\label{eq:current}
I_\textrm{tot} = I_\textrm{1e}+I_\textrm{2e}\,.
\end{equation}
The first contribution is due to the sequential tunneling of single electrons,
\begin{subequations}\label{eq:seq_Is}
\begin{align}\nonumber
I_\textrm{1e} &= e\,\sum_\alpha\,P_0\Gamma^l_{0\to \alpha}-P_{2}\Gamma^l_{2\to
\alpha}\\
&\;\;+e\sum_\alpha\,P_\alpha\left(\Gamma^l_{\alpha\to 2}- \Gamma^l_{\alpha\to
0}\right)\,.
\end{align}
The second contribution is due to the sequential tunneling of pairs of electrons via Andreev reflection processes,
\begin{equation}
I_\textrm{2e} = 2e\,\left(P_0\Gamma^l_{0\to 2} - \,P_{2}\Gamma^l_{2\to 0}\right)\,.
\end{equation}
\end{subequations}
The fact that the transition rates $\Gamma^l$ through the left junction appear in Eqs.~\eqref{eq:seq_Is}, rather than $\Gamma^r$, is due to a choice and not essential. In the steady state, the current at the left and right junctions must be equal by current conservation. Hence, one may obtain the same answer using the transition rates through the right junction instead.

The method of rate equations just outlined allows one to compute the current due to sequential (incoherent) tunneling processes. At low bias, this is the dominant contribution to the current close to the degeneracy points in the energy spectrum. Away from the Coulomb peaks, where direct tunneling into the nanowire is not allowed by energy conservation, the conductance is dominated by coherent co-tunneling processes, which need to be computed separately. As we will see, these are particularly important to capture the voltage dependence of the Coulomb peak tails at $\Delta<E_c$ (see Sec.~\ref{sec:elastic}).

\section{Andreev tunneling regime}\label{sec:andreev}

We begin by studying the case $B=0$, which is characterized by the presence of time-reversal symmetry. As a consequence, the energy spectrum of the system is Kramers degenerate. Thus, it is convenient to introduce a composite label $n\tau$ for the quasiparticle states in the hybrid nanowire, in place of the generic label $\alpha$ used in the previous sections; the integer $n$ labels different orbitals while $\tau$ is a Kramers index. The BCS Hamiltonian of Eq.~\eqref{eq:h_bcs} takes the familiar form
\begin{equation}\label{eq:h_wire_zero_field}
H^{B=0}_\textrm{BCS} = \sum_{n\tau} \epsilon_n \gamma^\dagger_{n
\tau}\gamma_{n\tau}+\textrm{const}\,,
\end{equation}
with $\epsilon_n^2=\zeta^2_n + \Delta_0^2$ and $\zeta_n$ being the single-particle energy of the $n$-th orbital in the normal state. Both $\zeta_n$ and $\epsilon_n$ are doubly degenerate. The corresponding solutions $u_{n\tau}(\vt{r}_j,\sigma)$ and $v_{n\tau}(\vt{r}_j,\sigma)$ of the Bogoliubov-de Gennes equation have a spin structure, due to the presence of spin-orbit coupling, and are directly related to the wave functions $\phi_{n\tau}(\vt{r}_j,\sigma)$ of the system in the normal state (solutions of the Schr\"odinger equation). Due to time-reversal symmetry, the latter functions satisfy the constraint
\begin{equation}
\phi_{n\tau}(\vt{r}_j,\sigma)=\sigma\tau\phi^*_{n\bar\tau}(\vt{r}_j,-\sigma)\,,
\end{equation}
where with $\bar\tau$ we denote the time-reversed partner of $\tau$. In terms of $\phi_{n\tau}$, the solutions of the Bogoliubov-de Gennes equations are given by
\begin{subequations}
\begin{align}
u_{n\tau}(\vt{r}_j,\sigma)&=u_n\phi_{n\tau}(\vt{r}_j,\sigma)\,,\\
v_{n\tau}(\vt{r}_j,\sigma)&=\sigma\,v_n\phi_{n\tau}(\vt{r}_j,-\sigma)\,,
\end{align}
\end{subequations}
with $u^2_n = \tfrac{1}{2}(1+\zeta_n/\epsilon_n)$ and $v^2_n = \tfrac{1}{2}(1-\zeta_n/\epsilon_n)$. That such a direct relation exists between the eigenfunctions of the Bogoliubov-de Gennes and Schr\"odinger equations is a consequence of time-reversal symmetry.

 The electron field operator of Eq.~\eqref{eq:Psi} and the tunneling Hamiltonian of Eq.~\eqref{eq:h_tunn} become
\begin{equation}\label{eq:Psi_zero_field}
\Psi(\vt{r},\sigma)=\sum_{n\tau} u_n\phi_{n\tau}(\vt{r}_j,\sigma)\,\gamma_{n\tau} + \sigma v_n \phi^*_{n\tau}(\vt{r}_j,-\sigma)\,\gamma^\dagger_{n\tau}
\end{equation}
and
\begin{align}\label{eq:h_tunn_zero_field}\nonumber
H^{B=0}_\textrm{tunn} & = \sum_{j,np,\tau\sigma}W_j\left[\theta(\Delta_\textrm{Al}-\epsilon_n)+\sqrt{\nu/\nu_\textrm{Al}}\,
\theta(\epsilon_n-\Delta_\textrm{Al})\right]\times\\
\nonumber
&\,\times[u_n\,\phi^*_{n\tau}(\vt{r}_j,\sigma)\,\phi_p(\vt{r}_j)\,\gamma^\dagger_{n\tau}\,c_{p\sigma}+\\
&+\sigma\,v_n\,\phi_{n\tau}(\vt{r}_j,-\sigma)\,\phi_p(\vt{r}_j)\,\gamma_{n\tau}\,c_{p\sigma}]+\textrm{H.c.}
\end{align}
respectively. Note that states with energy $\epsilon>\Delta_\textrm{Al}$ are mainly localized in the superconducting shell, where spin-orbit interaction is very weak, and therefore for these states spin is a good quantum number. For their wave functions one may therefore write $\phi_{n\tau}(\vt{r}_j,\sigma)=\delta_{\tau\sigma}\phi_n(\vt{r}_j)$. In any case, the presence or absence of spin-rotation symmetry has no drastic consequences on Andreev reflection as long as time-reversal symmetry is preserved.

We are now ready for the calculation of the current \cite{hekking1993}.  We assume that $\Delta_0$ is large enough that we are well far away from the transition to the single-electron tunneling at $B=B^*$. This requires $(\Delta_0-E_c)/E_c\gg T/T_p$. In this case, there is no poisoning effect: all transitions involving the odd parity states can be neglected. By setting $\Gamma_{0\to \alpha}=\Gamma_{2\to \alpha}=0$ in Eq.~\eqref{eq:steady_state_full}, we obtain
\begin{subequations}
\begin{align}
P_\alpha & = 0\,,\\
P_0 & = \frac{\Gamma_{2\to 0}}{\Gamma_{2\to 0}+\Gamma_{0\to 2}}\,,\\
P_2 & = \frac{\Gamma_{0\to 2}}{\Gamma_{2\to 0}+\Gamma_{0\to 2}}\,.
\end{align}
\end{subequations}
Thus, the sequential current is due solely to the pair contribution, which reads as
\begin{equation}\label{eq:I_Andreev}
I_\textrm{2e} = 2e\,\frac{\Gamma^l_{0\to 2}\Gamma^r_{2\to 0}
-\Gamma^l_{2\to 0}\Gamma^r_{0\to 2}}{\Gamma_{2\to 0}+\Gamma_{0\to
2}}\,.
\end{equation}
The rates $\Gamma_{0\to 2}$ and $\Gamma_{2\to 0}$ are due to Andreev reflection processes, in which a Cooper pair is either added or subtracted from the superconductor-wire hybrid. Because the pairing in the proximitized nanowire is purely $s$-wave, the two incoming or outgoing electrons must have opposite spins. Assuming that the occupation probabilities of single-particle states in the leads follow the Fermi-Dirac distribution, we get the following expressions from Fermi's Golden Rule:
\begin{subequations}\label{eq:Andreev_transition_rates}
\begin{align}\nonumber
\Gamma^j_{0\to 2} & = \frac{2\pi}{\hbar}\,\sum_{p_1p_2}\,|A^{j,0\to 2}_{p_1p_2}|^2\,\delta(E_0-E_2+\xi_{p_1}+\xi_{p_2})\times\\
&\qquad\qquad\qquad\times f(\xi_{p_1}-\mu_j)\,f(\xi_{p_2}-\mu_j)\,,
\\\nonumber\\\nonumber
\Gamma^j_{2\to 0}& = \frac{2\pi}{\hbar}\,\sum_{p_1p_2}\,|A^{j,2\to 0}_{p_1p_2}|^2\,\delta(E_0-E_2+\xi_{p_1}+\xi_{p_2})\,\times\\
&\qquad\qquad\qquad\times f(\xi_{p_1}+\mu_j)\,f(\xi_{p_2}+\mu_j)\,.
\end{align}
\end{subequations}
Here, $A^{j,0\to 2}_{p_1p_2}$ and $A^{j,2\to 0}_{p_1p_2}$ are amplitudes for the Andreev reflection processes that either add ($0\to 2$) or subtract ($2\to 0$) a Cooper pair from the BCS condensate, while subtracting or adding a pair of electrons from single-particle states $\ket{p_1+}$ and $\ket{p_2-}$ in lead $j$. Furthermore, $f(x) = [1+\exp(x/T)]^{-1}$ is the Fermi-Dirac distribution, we use the abbreviation $E_N=E_c(N-n_g)^2$, and we choose the chemical potential in the two leads to be $\mu_l = eV$ and $\mu_r = 0$.

Andreev reflection is a two-step process involving an intermediate state with one quasiparticle in the superconductor-wire hybrid, and its amplitude can be computed in second order in perturbation theory in the tunneling Hamiltonian \eqref{eq:h_tunn}. The result is
\begin{align}\nonumber
A^{j,0\to 2}_{p_1p_2} &= W^2_j\,\sum_{n\tau}\,\left[\theta(\Delta_\textrm{Al}-\epsilon_n)+\sqrt{\nu/\nu_\textrm{Al}}\,
\theta(\epsilon_n-\Delta_\textrm{Al})\right]\,\times\\\nonumber
& \times\,u_n\,v_n\,|\phi_{n\tau}(\vt{r}_j,+)|^2\phi_{p_1}(\vt{r}_j)\phi_{p_2}(\vt{r}_j)\times \\ & \times \left(\frac{1}{E_0-E_1+\xi_{p_1}-\epsilon_n}+\frac{1}{E_1-E_0+\xi_{p_2}-\epsilon_n}\right)\,.\label{eq:Andreev_amplitude}
\end{align}
The two contributions to the amplitude between the round brackets are distinguished by the order with which the two electrons in the lead tunnel into the superconductor  [note that a minus sign due to Fermi statistics is compensated by the factor $\sigma=\pm$ in Eq.~\eqref{eq:h_tunn_zero_field}, so that the two contributions interfere constructively].

So far our calculation applies, in fact, to any value of the ratio $\Delta/E_c$. At this point we make two simplifications. First, we may replace the energy difference $E_0-E_1=E_c(2n_g-1)$ with $E_c$, since we are mainly interested in the vicinity charge degeneracy point with $n_g=1$, and the residual dependence of the amplitude on $n_g$ would be weak. Second, under the condition $(\Delta_0 - E_c)/\Delta_0 \gg T/T_p$ we may neglect the energies $\xi_{p_1}$ and $\xi_{p_2}$ in the denominator, which are naturally limited by temperature. Hence we obtain
\begin{align}\nonumber
A^{j,0\to 2}_{p_1p_2} = W^2_j\sum_{n\tau}&\left[\theta(\Delta_\textrm{Al}-\epsilon_n)+\sqrt{\nu/\nu_\textrm{Al}}\theta(\epsilon_n-\Delta_\textrm{Al})\right]\times\\
\times&\,\frac{\Delta_0}{\epsilon_n}\,\frac{|\phi_{n\tau}(\vt{r}_j,+)|^2\phi_{p_1}(\vt{r}_j)\phi_{p_2}(\vt{r}_j)}{E_c-\epsilon_n}\,.
\end{align}
We now have to square the amplitude, which is a sum over many positive contributions $\propto |\phi_{n\tau}(\vt{r}_j,+)|^2$, making the fluctuations of $A^{j,0\to 2}_{p_1p_2}$ negligible. By performing the sum over $n$ in the continuum limit, one arrives at the expression
\begin{equation}\label{eq:Andreev_amp_squared}
\abs{A^{j,0\to 2}_{p_1p_2}}^2 = \frac{g_j^2\,\delta^2_j}{(2\pi)^4}\frac{16\Delta_0^2}{\Delta_0^2-E_c^2}\arctan^2\,\sqrt{\frac{\Delta_0+E_c}{\Delta_0-E_c}}\,,
\end{equation}
with $\delta_j$ the level spacing in the lead $j$. More precisely, the above equation may be interpreted as an average value of $\abs{A^{j,0\to 2}_{p_1p_2}}^2$, which for instance can be obtained by sampling the wave functions $\phi_{n\tau}(\vt{r}_j,\sigma)$ from the Gaussian symplectic ensemble or, for those states with spin-rotation symmetry, the Gaussian orthogonal ensemble \cite{mehta2004random}. Inserting Eq.~\eqref{eq:Andreev_amp_squared} in Eq.~\eqref{eq:Andreev_transition_rates} and performing the summation over states in the leads, we obtain
\begin{align}\nonumber
\Gamma^j_{0\to 2} &=\frac{2\pi}{\hbar}\frac{g^2_j}{(2\pi)^4}\,\frac{E_2-E_0-2\mu_j}{\exp[(E_2-E_0-2\mu_j)/T]-1}\\
&\times\frac{16\Delta_0^2}{\Delta_0^2-E_c^2}\arctan^2\,\sqrt{\frac{\Delta_0+E_c}{\Delta_0-E_c}}\,.
\end{align}
The expression for the other transition rate $\Gamma^j_{2\to 0}$ can be obtained by sending $\mu_j\to -\mu_j$ and $E_2-E_0\to E_0-E_2$\,. Inserting the transition rates in Eq.~\eqref{eq:I_Andreev}, we get the following expression for the zero-bias differential conductance at $B=0$ \cite{hekking1993},
\begin{align}\nonumber
G_\textrm{2e}&=\frac{2e^2}{h}\,\frac{g^2_lg^2_r}{g^2_l+g^2_r}\,\frac{4E_c(1-n_g)/T}{\sinh[4E_c(1-n_g)/T]}\frac{\Delta_0^2}{\Delta_0^2-E_c^2}\\
&\times\,\frac{4}{\pi^2}\arctan^2\sqrt{\frac{\Delta_0+E_c}{\Delta_0-E_c}}\,.\label{eq:G_2e}
\end{align}
The conductance exhibits a symmetric peak around the point $n_g=1$. The peak height is temperature independent, while the peak width is proportional to $T$.

A weak magnetic field will not affect dramatically the final result of Eq.~\eqref{eq:G_2e} as long as the corresponding Zeeman energy remains small compared to $\Delta_0$. In this case, the singlet condensate is only weakly affected by the breaking of time-reversal symmetry and the energies of the virtual states in Eq.~\eqref{eq:Andreev_amplitude} are split by a small amount $\sim g \mu_B B$. Note, however, that a more drastic effect of the magnetic field should be observed if the leads are comprised from long segments of a single-channel wire with strong spin-orbit interaction. In this case, as already mentioned \cite{Note1}, Zeeman splitting removes one of the propagating modes. The Andreev reflection for the electrons impinging on the junction via the remaining single propagating mode is suppressed at the Fermi energy \cite{beri2009,wimmer2011}, and we expect conductance suppression as long as $\max(eV,T) \lesssim g \mu_B B$.

\section{Single electron tunneling regime}\label{sec:single_electron}

Let us now consider the regime in which the magnetic field is large, such that $\Delta(B)<E_c$, and in which the nanowire is approaching the topological phase transition, $B\lesssim B_c$ with $T\ll\Delta(B)\ll \Delta_0$. To characterize the low-energy spectrum of the proximitized nanowire in this regime, we may use the toy-model of a single-channel nanowire \cite{oreg2010,lutchyn2010}, which for a system of infinite length predicts a gap closing at zero momentum. For a wire of length $L$, the low-energy spectrum approximately is
\begin{equation}\label{eq:low_energy_spectrum}
\epsilon_n = \sqrt{\Delta^2 + \delta^2 (n+1/2)^2}\,,\;\;n=0,1,2,\dots
\end{equation}
Both time-reversal and spin-rotation symmetries are broken, so there are no good quantum numbers beside the orbital index $n$, and no degeneracies in the spectrum. The $1/2$ offset in Eq.~\eqref{eq:low_energy_spectrum} is due to the confinement energy for the plane wave states \cite{mishmash2016}.

Using the simple model of Refs.~\cite{lutchyn2010,oreg2010}, we have checked numerically that Eq.~\eqref{eq:low_energy_spectrum} is a very good approximation of the low-lying states of a finite size nanowire, at least as long as one can neglect the branches of the energy spectrum at large momentum $\abs{k}\sim k_F$. The pairing gap for these branches remains close to $\Delta_0$ for a strongly spin-orbit coupled wire where the spin-orbit energy $E_\textrm{so}=m\alpha^2$ dominates the Zeeman energy ($m$ is the effective mass in the semiconducting nanowire). Under this condition, there are $\sim\Delta_0/\delta$ states whose energies are well approximated by Eq.~\eqref{eq:low_energy_spectrum}. The number of states contributing to transport is at the same time limited by temperature, and of the order of $\sqrt{T\Delta}/\delta < \Delta_0/\delta$. We conclude that it is indeed sufficient to focus on this region of the spectrum.

As mentioned in the Introduction, the level spacing $\delta$ in Eq.~\eqref{eq:low_energy_spectrum} depends on the strength of the proximity effect. In Appendix~\ref{app:level_spacing}, we show that $\delta$ can be estimated in terms of measurable parameters of the hybrid system as
\begin{equation}\label{eq:level_spacing}
\delta = Z_0\,\frac{\pi \alpha}{L}\,,
\end{equation}
with
\begin{equation}\label{eq:Z0}
Z_0=\frac{\Delta_\textrm{Al}}{\Delta_\textrm{Al}+\Delta_0\sqrt{(\Delta_\textrm{Al}+\Delta_0)/(\Delta_\textrm{Al}-\Delta_0)}}\,
\end{equation}
The estimate \eqref{eq:level_spacing} assumes that the chemical potential in the nanowire is situated in the middle of the Zeeman gap, which is the optimal value. Equation~\eqref{eq:level_spacing} quantifies the intuitive fact that for a strongly proximitized nanowire ($\Delta_0\to \Delta_\textrm{Al}$, $Z_0\to 0$), the level spacing is renormalized downwards due to the hybridization with states in Al (see Ref.~\cite{pientka2015} for an analysis of the same effect). On the other hand, for weak or vanishing proximity ($\Delta_0\to 0$, $Z_0\to 1$), the level spacing tends to the inverse dwell time $\pi \alpha/L$ of an electron propagating ballistically through the nanowire. For instance, Eq.~\eqref{eq:level_spacing} gives $\delta\approx 5.6\,\mu$eV for $L=2\,\mu$m, $\alpha=10\,\mu$eV$\cdot\mu$m, $\Delta_0=180\,\mu$eV and $\Delta_\textrm{Al}=130\,\mu$eV. These are the values used in Figs.~\ref{fig:g_seq} and \ref{fig:elastic_cond}.

The effective low-energy Hamiltonian and electron field operator now read just like Eqs.~\eqref{eq:h_wire}, \eqref{eq:h_bcs} and \eqref{eq:Psi}, respectively, but with the label $\alpha$ replaced by integer $n$, and $\epsilon_n$ specified in Eq.~\eqref{eq:low_energy_spectrum}. Note that, in the tunneling Hamiltonian, we limit the summation to the low-lying states described by~\eqref{eq:low_energy_spectrum} . We are now ready for the evaluation of the conductance. We split the calculation in two parts: in the next subsection we compute the sequential tunneling contribution which determines the peak value of the conductance, and afterwards we focus on the elastic co-tunneling contribution.

\subsection{Sequential tunneling}

For the sequential tunneling contribution we start again from the steady state solution of Sec.~\ref{sec:rate_eqs}. Since now $\Delta<E_c$, we may neglect all transition rates which bring the wire into the $\ket{2}$ state. Setting $\Gamma_{0\to 2}=0$ and $\Gamma_{\alpha \to 2}=0$ in Eq.~\eqref{eq:steady_state_full}, we obtain (replacing the label $\alpha$ with $n$) the following steady-state occupation probabilities,
\begin{subequations}
\begin{align}
P_0 & = \left(1 + \sum_n \frac{\Gamma_{0\to n}}{\Gamma_{n \to
0}}\right)^{-1}\,,\\
P_n& = \frac{\Gamma_{0\to n}}{\Gamma_{n \to 0}}\, \left(1 +
\sum_n \frac{\Gamma_{0\to n}}{\Gamma_{n \to 0}}\right)^{-1}\,,\\
P_{2} & = 0\,.
\end{align}
\end{subequations}
Replacing the above expressions in Eq.~\eqref{eq:seq_Is}, we see that current due to sequential tunneling of electrons is given by
\begin{align}\nonumber
I_\textrm{e}  = e&\left(1 + \sum_n \frac{\Gamma_{0\to n}}{\Gamma_{n\to 0}}\right)^{-1}\times \\
& \quad\times\sum_n \left(\Gamma^l_{0\to n} - \Gamma^l_{n\to 0}
\frac{\Gamma_{0\to n}}{\Gamma_{n\to 0}}\right)\,.\label{eq:I_e}
\end{align}
We now need to compute the transition rates $\Gamma_{0\to n}$ and $\Gamma_{n\to0}$ which describe the tunneling of a single charge between the contacts and the wire. This is a first-order process which may involve any of the states in the leads, and again using Fermi's Golden Rule one finds
\begin{align}\nonumber
\Gamma^j_{0\to n}&= \frac{2\pi}{\hbar}\sum_{p\sigma}W_j^2\,|\phi_p(\vt{r}_j)|^2\,|u_n(\vt{r}_j,\sigma)|^2\,\\\nonumber
&\qquad\qquad\times\delta(E_0-E_1+\xi_p-\epsilon_n) f(\xi_p-\mu_j)\,,\\\nonumber
\Gamma^j_{n \to 0}&=\frac{2\pi}{\hbar}\sum_{p\sigma}W_j^2\,|\phi_p(\vt{r}_j)|^2\,|u_n(\vt{r}_j,\sigma)|^2\\
&\qquad\qquad\times\delta(E_0-E_1+\xi_p-\epsilon_n)\,[1-f(\xi_p-\mu_j)]\,
\label{eq:gamma_single_el}
\end{align}
with $\mu_l=eV$, $\mu_r=0$. Factor $f(x)$ or $1-f(x)$ appears in the equations above depending on whether the transfer of charge subtracts or adds an electron in the single-particle state $\ket{p\sigma}$ of the lead.

In order to proceed, we need to know the values of $|u_n(\vt{r}_j,\sigma)|^2$. From the normalization condition we can write $|u_n(\vt{r}_j,\sigma)|^2\sim Z_0/L$, but computing the missing prefactor is a non-trivial task, not even in the clean limit, since contrary to time-reversal symmetric case treated in Sec.~\ref{sec:andreev}, the energy and spatial dependence of $u_n(\vt{r}_j,\sigma)$ can not be inferred easily from the eigenstates of the system in the normal state. However, even without entering into microscopic details, we know that the electron and hole parts $(u_n, v_n)$ of a solution of the Bogoliubov-de Gennes equations with energy $\epsilon_n$ close to $\Delta$ have almost equal weight. Hence, using the completeness relation in spin space, we obtain
\begin{equation}
W_j^2\,|\phi_p(\vt{r}_j)|^2\,\sum_\sigma |u_n(\vt{r}_j,\sigma)|^2\simeq \frac{g_j\,\delta_j \delta}{2\,(2\pi)^2}\,,
\end{equation}
independent of $n$ to leading order in $\delta$. This leads to
\begin{align}\nonumber
\Gamma^j_{0\to n}&= \frac{g_j\delta}{4\pi\hbar}\,f(E_1-E_0+\epsilon_n-\mu_j)\,,\\
\Gamma^j_{n \to 0}&=\frac{g_j\delta}{4\pi\hbar}\,[1-f(E_1-E_0+\epsilon_n-\mu_j)]\,.
\end{align}

\begin{figure}[t!]
\begin{center}
\includegraphics[width=\columnwidth]{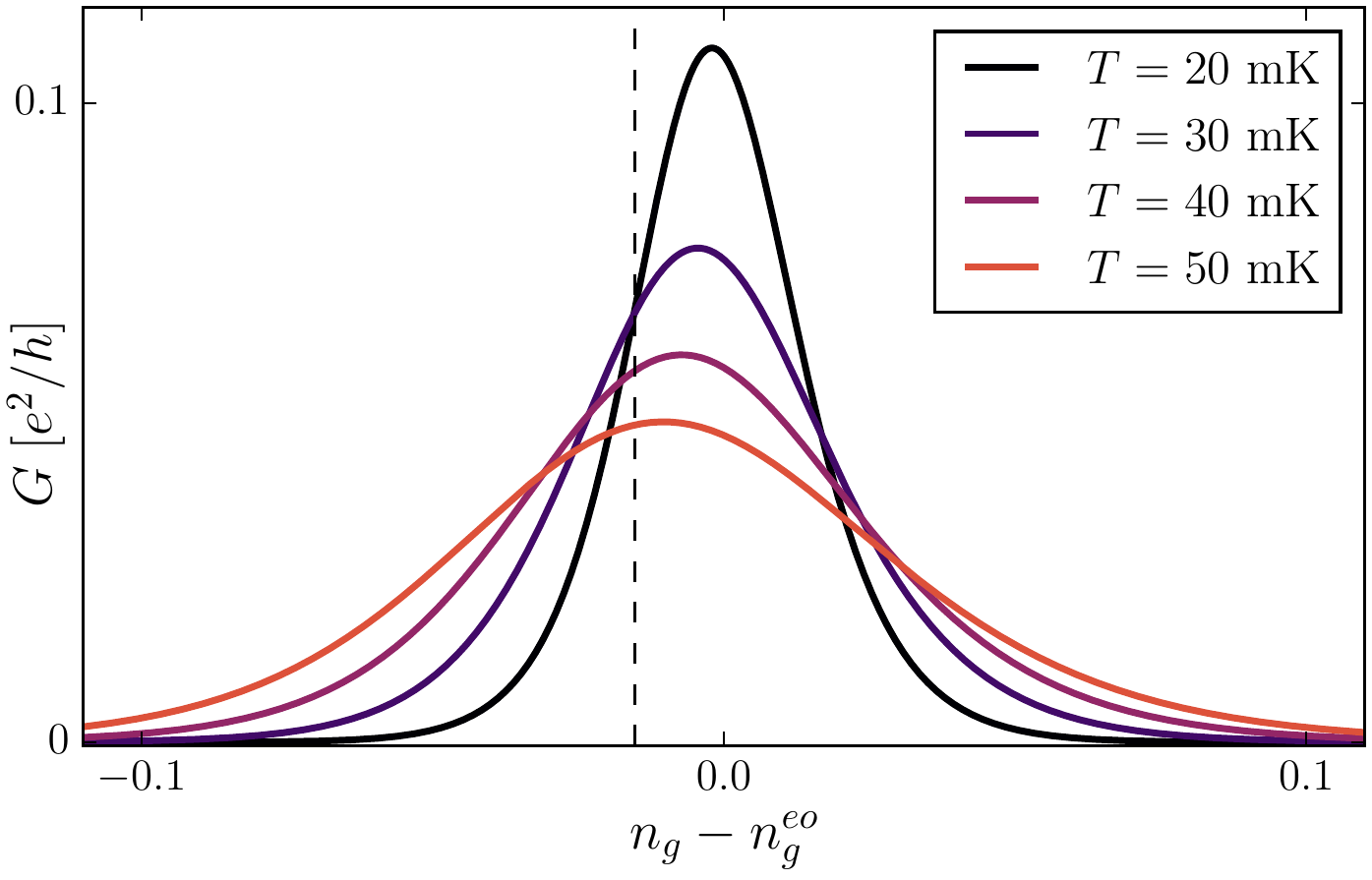}
\caption{Plot of the conductance peak due to single electron tunneling at $\Delta<E_c$ and for different temperatures, obtained by a numerical summation of Eq.~\eqref{eq:sequential_peak}. We have used the following parameters: $\Delta_\textrm{Al}=180\,\mu$eV, $\Delta_0=130\,\mu$eV, $\Delta=80\,\mu$eV, $E_c=100\,\mu$eV, $\alpha=0.01$ eV$\cdot$nm and $L=2\,\mu$m. According to Eq.~\eqref{eq:level_spacing}, for these parameters the level spacing is $\delta\approx 5.6\,\mu$eV. The summation in Eq.~\eqref{eq:sequential_peak} was truncated after $[\Delta_0/\delta]=23$ terms. The thin dashed vertical line marks the value of $\eta_\textrm{peak}$ for $T=50$ mK, estimated from Eq.~\eqref{eq:seq_peak_pos}. For the curves in the figure, the missing numerical prefactor in Eq.~\eqref{eq:g_seq_peak} is $\approx 0.25$. \label{fig:g_seq}}
\end{center}
\end{figure}

After inserting this result in Eq.~\eqref{eq:I_e} and after a tedious but straightforward application of the chain rule, we obtain for the differential conductance at zero bias voltage the following expression
\begin{align}\nonumber
G_\textrm{e} &=
\frac{e^2}{2h}\frac{g_lg_r}{g_l+g_r}\frac{\delta}{T}\,\frac{1}{1 + \sum_n \exp[(\Delta-\epsilon_n+2\eta E_c)/T]}\times\\
&\qquad\times\sum_{n}\,\frac{1}{1+\,\exp[(\epsilon_n-\Delta-2\eta E_c)/T]}\,,\label{eq:sequential_peak}
\end{align}
where $\eta = n_g - n_g^\textrm{eo}$ is a parameter which measures the vicinity to the even-odd charge degeneracy point. The sum in the denominator of Eq.~\eqref{eq:sequential_peak} can be computed in the continuum limit, with the result $\sum_n \exp[(\Delta-\epsilon_n)/T]=\sqrt{T/4\Delta}\,\exp(\Delta/T_p)$ for $T\ll\Delta$. On the other hand, the remaining sum in Eq.~\eqref{eq:sequential_peak} can not be performed analytically, but by studying the dominant contribution to the corresponding integral we obtain the following estimates. The conductance exhibits peaks of height
\begin{equation}\label{eq:g_seq_peak}
G_e^\textrm{peak}\sim \frac{e^2}{h}\,\frac{g_lg_r}{g_l+g_r}\,\frac{\delta}{T}\,,
\end{equation}
with a numerical prefactor of order one. The finite-temperature peak position $\eta_\textrm{peak}(T)$ is shifted from its $T=0$ value $\eta_{\rm peak}(0)=0$,
\begin{equation}\label{eq:seq_peak_pos}
\eta_\textrm{peak}(T) \approx -\frac{T}{4E_c}\,\left[\frac{\Delta}{T_p} + \frac{1}{2}\,\ln\left(\frac{T}{4\Delta}\right)\right]\,.
\end{equation}
The peak width is of the order of $|\eta_\textrm{peak}(T)|$. The ``tails'' of the peak are exponentially small, $G_e\!\sim\!\sqrt{\Delta/T}\exp(-2\abs{\eta}E_c/T)$ for $\abs{\eta}\gg\abs{\eta_\textrm{peak}}$ .

In Fig.~\ref{fig:g_seq} we plot the conductance peak obtained via a numerical summation of Eq.~\eqref{eq:sequential_peak}, for different values of the temperature. It shows how the peak position shifts more towards the left of the charge degeneracy point $\eta=0$ (i.e, $n_g=n_g^\textrm{eo}$) with increasing $T$. The temperature increase makes the difference in the size of even and odd Coulomb valleys as a function of gate voltage less and less pronounced, a consequence of poisoning of the proximitized nanowire. Analysis of Eq.~\eqref{eq:g_seq_peak} also indicates that the peak has a width proportional to $\eta_\textrm{peak}$, and that it is slightly asymmetric with larger conductance on the odd side ($\eta>\eta_\textrm{peak}$). This asymmetry, however, is hardly seen in the thermal tails of the peak, because of their exponential smallness at low temperatures. At $|\eta|\gtrsim T/\Delta$, the temperature-independent elastic co-tunneling contribution dominates the conductance. As we show next, it brings a conductance asymmetry of opposite sign with respect to the peak position, and yields a larger conductance on the even side of the peak ($\eta<\eta_\textrm{peak}$).

\subsection{Elastic co-tunneling}\label{sec:elastic}

\begin{figure}[t!]
\begin{center}
\includegraphics[width=\columnwidth]{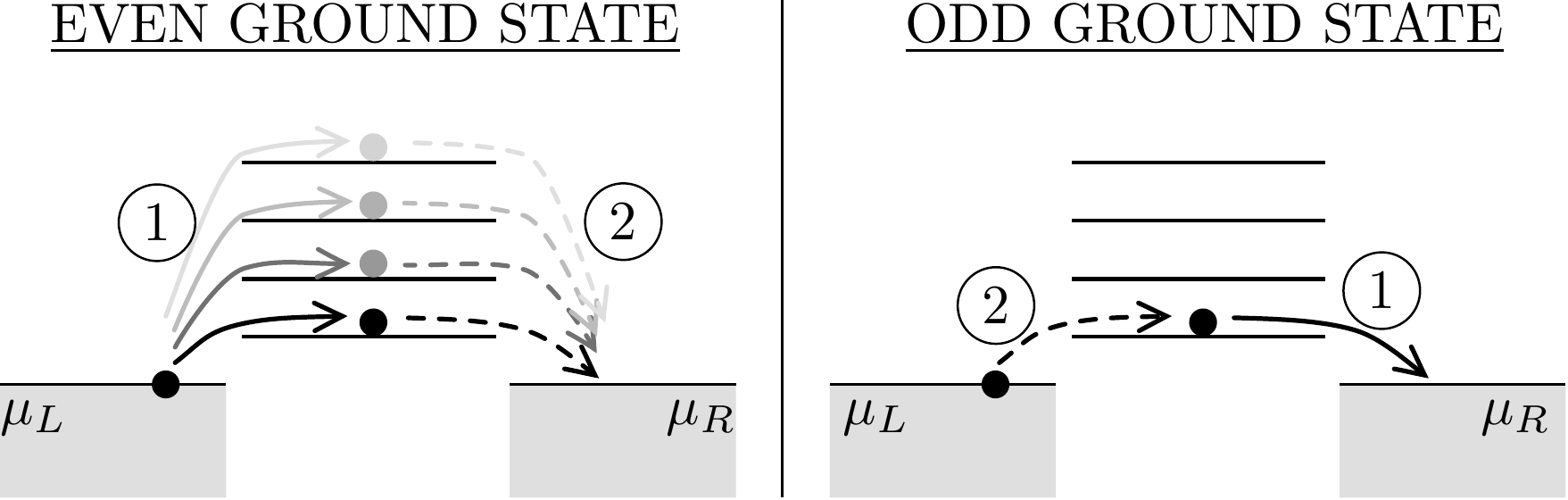}
\caption{Illustration of the even-odd asymmetry of elastic co-tunneling, which is a two-step process involving a sequence of two coherent tunneling events. Their order is indicated in the figure by the numbers and differs for the even and odd electron numbers in the ground state. In the even case (left panel), initially there are no quasiparticles in the proximitized nanowire. An electron may then tunnel from the left lead (step 1), occupy virtually any of the low-lying states of the nanowire, and then tunnel out to the right lead (step 2). The different intermediate states are indicated by the different gray arrows. For odd ground state parity (right panel), instead, one quasiparticle initially occupies the lowest energy level of the proximitized nanowire. The elastic co-tunneling therefore is completed following a reversed order: first the additional quasiparticle tunnels out to the right lead, and then it is replaced by an electron from the left lead. This is the only resonant contribution to the amplitude close to the charge degeneracy point: contributions involving other intermediate states are blockaded due to the high charging energy cost associated with the contemporary presence of two excess electrons in the nanowire, or with the breaking of a Cooper pair. \label{fig:diagram_elastic_tunn}}
\end{center}
\end{figure}

The term elastic co-tunneling refers here to a coherent transfer of electrons between the leads via a virtual state in the wire. For a generic superconducting island with $\Delta<E_c$, it was first studied by Averin and Nazarov~\cite{averin1992}, who found that $G_\textrm{el} \sim (e^2/h) g_l g_r \delta/\Delta$. They did not focus on its dependence on the gate voltage $n_g$, which is indeed very weak far away from the degeneracy points. Our motivation to revisit this transport process in detail is the observation of a large asymmetry in the conductance peak in the regime $\Delta<E_c$ \cite{higginbotham2015}, with the conductance on the even side of the peak ($n_g<n_g^\textrm{eo}$) being larger than on the odd side ($n_g>n_g^\textrm{eo}$). The asymmetry was more pronounced at lower temperatures and was observed in relatively short wires. Here we argue that a possible explanation of this effect lies precisely in the elastic contribution $G_\textrm{el}$ to the conductance. We extend the analysis of Ref.~\cite{averin1992} and find that $G_\textrm{el}$ strongly enhances one side of the peak, $n_g<n_g^{eo}$, the one that corresponds to the even electron number in the ground state. The enhancement is due to a large number of nearly-resonant contributions to the tunneling amplitude. On the odd side, on the contrary, all these contributions to the elastic co-tunneling amplitude are suppressed due to the presence of an unpaired quasiparticle, as illustrated in Fig.~\ref{fig:diagram_elastic_tunn}. We now present our quantitative results which support aforementioned qualitative considerations.

In an elastic co-tunneling process, an electron with energy $\xi_p$ and spin $\sigma_1$ is transferred to a state with spin $\sigma_2$ and same energy in the right lead (or vice versa). The process leaves behind no quasiparticle excitations in the proximitized nanowire. The total current can be obtained using the Fermi's Golden Rule, as
\begin{align}\label{eq:I_el}\nonumber
I_\textrm{el} =
\frac{2\pi e}{\delta_r\hbar}\sum_{p,\sigma_1\sigma_2}&\Big(P_\textrm{0}
\abs{A^\textrm{el}_0}^2 + \sum_n P_n\,\abs{A^\textrm{el}_n}^2
+ P_2 \abs{A^\textrm{el}_2}^2\Big)\\&\;
\times[f(\xi_p-eV)-f(\xi_p)]\,.
\end{align}
We have used the energy conservation to eliminate a summation over states in the right lead. The tunneling process is characterized by an amplitude $A^\textrm{el}_i$ which depends on the initial state $\ket{i}$ of the nanowire (we omit the explicit dependence of the amplitude on $p$, $\sigma_1$ and $\sigma_2$). The contribution of each amplitude must be weighted by the probability $P_i$ for the wire to be in state $\ket{i}$.

We are interested in the elastic contribution, Eq.~\eqref{eq:I_el}, outside the domain of thermal broadening of the conductance peak, see Eq.~\eqref{eq:seq_peak_pos}. Therefore, we may set $T=0$ in the evaluation of $G_\textrm{el}$.
At $T=0$, the occupation probabilities $P_0, P_n, P_2$ in Eq.~\eqref{eq:I_el} are simply determined by the ground state for a given value of $n_g$. We then simplify Eq.~\eqref{eq:I_el}
\begin{equation}\label{eq:g_el_initial}
G_\textrm{el} = \frac{2\pi e^2}{\hbar}\frac{1}{\delta_l\delta_r}\times
\begin{dcases}
\sum_{\sigma_1\sigma_2}\abs{A^\textrm{el}_0}^2 & n_g<n_g^\textrm{eo}\,,\\
\sum_{\sigma_1\sigma_2}\abs{A^\textrm{el}_{n=0}}^2 & n_g>n_g^\textrm{eo}\,,
\end{dcases}
\end{equation}
where the amplitudes $A^\textrm{el}_0$ and $A^\textrm{el}_{n=0}$ are for an incoming electron at the Fermi level. They are obtained in second order in the tunneling Hamiltonian and involve a sum over intermediate states. In computing amplitude $A^\textrm{el}_0$ we assume the nanowire is initially in even state with no quasiparticles and we obtain
\begin{subequations}\label{eq:elastic_cot_sums}
\begin{equation}
A^\textrm{el}_0 =
W_lW_r\,\phi_{p_1}(\vt{r}_l)\phi^*_{p_2}(\vt{r}_r)\sum_n\,\frac{u_n^*(\vt{r}_l,\sigma_1)u_n(\vt{r}_r,\sigma_2)}{E_c(2n_g-1)-\epsilon_n}\,.\label{eq:elastic_cot_sums_even}
\end{equation}
The sum here corresponds to a manifold of states with an extra electron occupying one of the quasiparticle states in the proximitized wire. On the other hand, when computing $A^\textrm{el}_{n=0}$ we assume that a quasiparticle is present in the lowest energy level of the spectrum of Eq.~\eqref{eq:low_energy_spectrum}, and we obtain
\begin{align}\label{eq:elastic_cot_sums_odd}
&A^\textrm{el}_{n=0} = W_lW_r\,\phi_{p_1}(\vt{r}_l)\phi^*_{p_2}(\vt{r}_r)\,\\
&\times\left[\frac{u_0^*(\vt{r}_l,\sigma_1)u_0(\vt{r}_r,\sigma_2)}{E_c(2n_g-1)-\epsilon_0} - \sum_{n\neq 0}\frac{v_n(\vt{r}_l,\sigma_1)v^*_n(\vt{r}_r,\sigma_2)}{E_c(2n_g-1)+\epsilon_{n}}\right]\, \nonumber.
\end{align}
\end{subequations}
The first term here corresponds to a virtual intermediate state in which the unpaired electron initially present in the ground state tunnels out from the nanowire. The sum reflects virtual states formed by breaking a Cooper pair and extracting one of the constituent electrons from the nanowire; the energy of these states is larger than $2\Delta$ at any value of $n_g$. In writing both amplitudes, we neglected contributions involving intermediate states with charge different from $N$ or $N+1$. These would appear at the same order in the tunneling matrix elements, but involve intermediate states with an energy larger by an amount at least $E_c$.

The energies of the intermediate states involved respectively in Eqs.~\eqref{eq:elastic_cot_sums_even} and \eqref{eq:elastic_cot_sums_odd} differ drastically from each other when $n_g\to n_g^\textrm{eo}=(\Delta+E_c)/2E_c$, that is when $E_c(2n_g-1)\to \Delta$. By looking at the denominators in Eq.~\eqref{eq:elastic_cot_sums_even}, we see that the amplitude for the even states contains many contributions with a small denominator of order $\delta$. On the odd side, instead, there is only one such contribution, represented by the first term in Eq.~\eqref{eq:elastic_cot_sums_odd}, while all others have a denominator which is at least $2\Delta$.

\begin{figure}[t!]
\begin{center}
\includegraphics[width=\columnwidth]{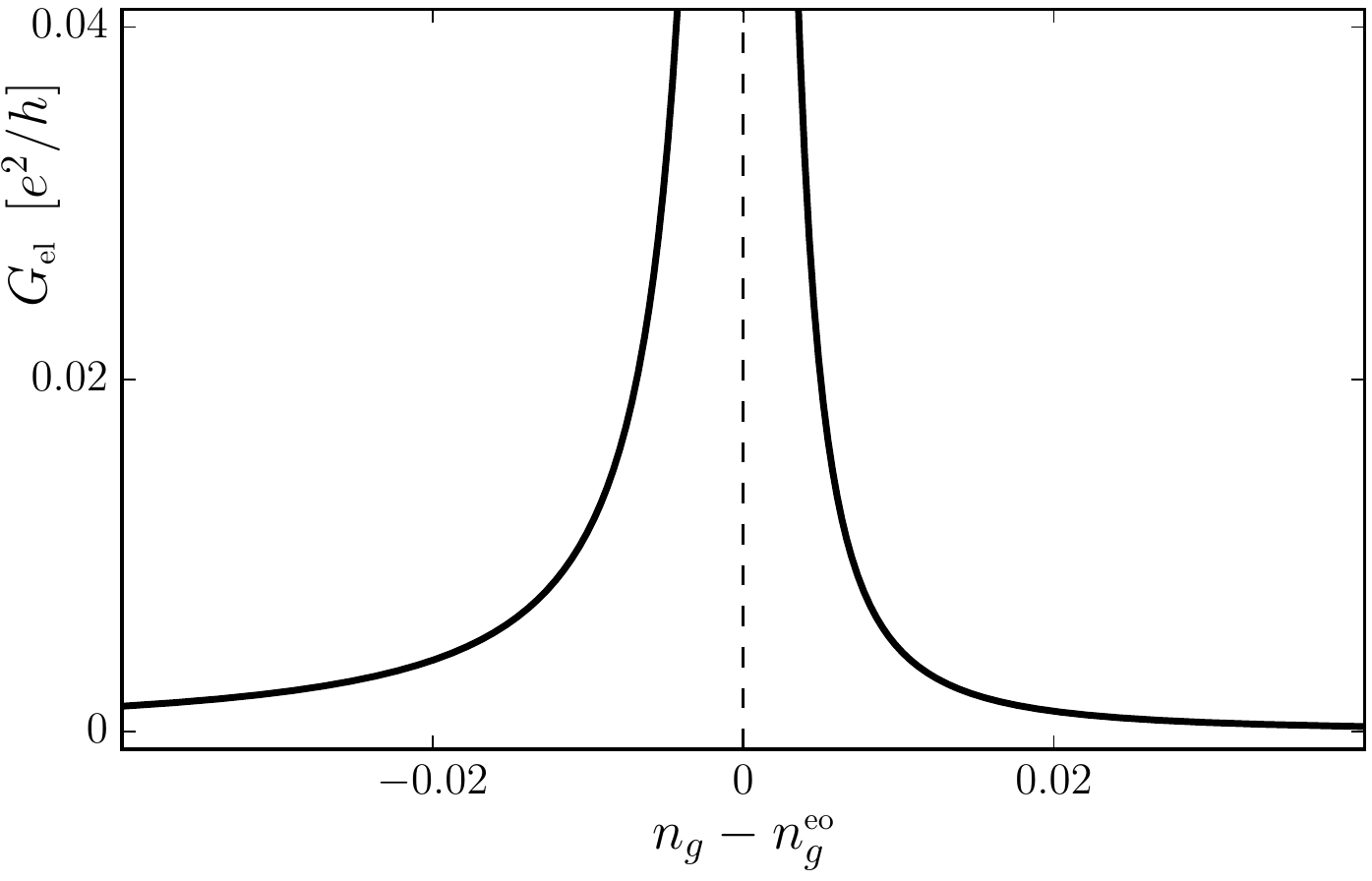}
\caption{Plot of the $T=0$ elastic co-tunneling conductance on both sides of the degeneracy point $n_g=n_g^\textrm{eo}$. The curve is obtained via a numerical summation of Eq.~\eqref{eq:co_tunneling_ampitudes_zero_T}, using the same parameters as in Fig.~\eqref{fig:g_seq}.\label{fig:elastic_cond}}
\end{center}
\end{figure}

In order to proceed with the calculation, we need to square the amplitudes in Eqs.~\eqref{eq:elastic_cot_sums}. In doing so, we use the fact that the phases of the wave functions are different for different states, resulting in an effective cancellation of the cross-terms appearing upon squaring the sums present in Eqs.~\eqref{eq:elastic_cot_sums}. To leading order in $\delta$, we may therefore replace the absolute square of the sum with the sum of squares \cite{averin1990}. For instance, when inserting Eq.~\eqref{eq:elastic_cot_sums_even} in Eq.~\eqref{eq:g_el_initial} the crucial part of the calculation goes as follows:
\begin{align}\nonumber
&\sum_{\sigma_1\sigma_2}\Bigl|\sum_n\,\frac{u_n^*(\vt{r}_l,\sigma_1)u_n(\vt{r}_r,\sigma_2)}{E_c(2n_g-1)-\epsilon_n}\Bigr|^2\simeq \\\nonumber
&\quad\simeq\sum_{\sigma_1\sigma_2}\sum_n \frac{|u_n(\vt{r}_l,\sigma_1)|^2\,|u_n(\vt{r}_r,\sigma_2)|^2}{E_c(2n_g-1)-\epsilon_n}\simeq\\
&\quad\simeq\frac{Z_0^2}{4L^2}\sum_n\frac{1}{[E_c(2n_g-1)-\epsilon_n]^2}\,.
\end{align}
We recall that the factor $Z_0^2$ appears from the normalization of the wave function. In going from the second line to the third line above, we have again used the completeness of the basis in spin space, as well as the fact that for $\epsilon_n$ close to $\Delta$ the electron and hole parts of the quasiparticle wave functions have equal weight.
Note that while this procedure essentially allows us to obtain an average value of the conductance, we expect substantial fluctuations between different Coulomb blockade valleys \cite{aleiner1996}. For the average value of the conductance, we obtain
\begin{subequations}\label{eq:co_tunneling_ampitudes_zero_T}
\begin{equation}
G_\textrm{el} = \frac{e^2}{h}\frac{g_l g_r \delta^2}{(2\pi)^2}
\frac{1}{4}\sum_n\frac{1}{[E_c(2n_g-1)-\epsilon_n]^2} \quad \textrm{if}\;
n_g<n_g^\textrm{eo}\,,
\end{equation}
and
\begin{align}\nonumber
G_\textrm{el} &= \frac{e^2}{h}\frac{g_l g_r \delta^2}{(2\pi)^2} \,\frac{1}{4}\left
\{\frac{1}{[E_c(2n_g-1)-\epsilon_0]^2}+\right.\\
&\left. \qquad+\sum_{n>0} \frac{1}{[E_c(2n_g-1)+\epsilon_n]^2}\right\}\quad
\textrm{if}\; n_g>n_g^\textrm{eo}\,.
\end{align}
\end{subequations}
These sums can be performed numerically as illustrated in Fig.~\ref{fig:elastic_cond}. They can also be evaluated analytically in the continuum limit, applicable at $\abs{\eta}\gtrsim \delta^2/\Delta E_c$, which allows us to obtain in the linear order in $\delta$ the following asymptotic behavior close to the charge degeneracy point,
\begin{equation}\label{eq:elastic_cotunn_asymptotes}
G_\textrm{el} \sim \frac{e^2}{h}\frac{g_l g_r}{4\,(2\pi)^2}\times
\begin{dcases}
\frac{\delta}{2E_c}\left(\frac{\Delta}{2E_c}\right)^{1/2}\frac{1}{|\eta|^{3/2}}, & \eta \to 0^-\,,\\
\frac{2}{3}\,\frac{\delta}{\Delta}\,& \eta \to 0^+\,,
\end{dcases}
\end{equation}
where, we recall, $\eta=n_g-n_g^\textrm{eo}$. The single diverging contribution present in Eq.~\eqref{eq:co_tunneling_ampitudes_zero_T} for $n_g>n_g^\textrm{eo}$ adds to the conductance on the odd side of the peak a higher-order in $\delta$ term, $\sim (e^2/h)(g_lg_r/4\pi^2)\,\delta^2/(16E_c^2\eta^2)$, which can only compensate for the asymmetry in a narrow interval $\abs{\eta}\lesssim \delta^2/(\Delta E_c)$.

The divergence at $n_g = n_g^\textrm{eo}$  is, of course, not physical. At finite temperature, it can be removed by the regularization procedure outlined in Ref.~\cite{turek2002}. The regularization only affects the result in the vicinity of the transition point, for $\abs{\eta} \lesssim (g_l+g_r)\,\delta/E_c$, and therefore does not affect the conclusion about the asymmetry of the conductance peak indicated by Eq.~\eqref{eq:elastic_cotunn_asymptotes} as long as $T\gtrsim (g_l+g_r)\,\delta$.

Finally, we compare the elastic co-tunneling and the sequential tunneling contributions to the conductance. By equating the even-side asymptote of Eq.~\eqref{eq:elastic_cotunn_asymptotes} with the activated tails of Eq.~\eqref{eq:sequential_peak}, we see that the elastic co-tunneling dominates over the thermal tail and thus defines the conductance asymmetry with respect to $\eta$ at $|\eta|\gtrsim |\eta^*|$, with
\begin{equation}
\eta^* \approx - \frac{T}{4 E_c}\,\ln\,\left[\frac{E_c^3}{\delta^2(g_l+g_r)^2T}\right]\,.
\end{equation}
As expected, the elastic co-tunneling is enhanced at low temperatures. Increasing the level spacing $\delta$, the conductances of the point contacts, or the charging energy $E_c$ also enhances the relative weight of the elastic co-tunneling process to the total conductance.

The asymmetry of the Coulomb blockade peaks in the single-electron tunneling regime is due to the different nature of the excitations spectra in the two charge states brought to resonance: the spectral gaps, $\sim\Delta$ in the even state and $\sim\delta^2/\Delta$ in the odd one, are vastly different. The asymmetry of the peaks may complicate finding the energy of the spatially-quantized quasiparticle levels from the position of the peaks (this technique was widely used in the physics of semiconductor quantum dots~\cite{simmel1997,alhassid2000}, and may require to attain very low temperatures, $T\lesssim\delta^2/\Delta$. In that temperature range, we expect small shifts, $\sim\delta^2/(\Delta E_c)$ of the peak positions compared to the nominal ones, $n_g=n_g^\textrm{eo}$.

\section{Resonant tunneling through Majorana bound states}\label{sec:majorana}

Let us now move on to the case $B>B_c$. If the proximitized nanowire is in the topological phase, it will host two Majorana bound states close to the two point contacts. In the ideal case, the single-particle spectrum of the nanowire consists of a single zero-energy quasiparticle state separated by a gap $\Delta(B)$ from the quasi-continuum of extended states in the nanowire. The even-odd charge degeneracy point is now situated at $n_g^\textrm{eo}=1/2$, and similarly to the Coulomb blockade in a metallic island one expects conductance peaks with a periodicity in gate voltage corresponding to a single electron charge. In this situation, the leading mechanism for conduction is the resonant tunneling mediated by the pair of Majorana bound states \cite{fu2010,zazunov2011,hutzen2012}; away from the charge degeneracy point resonant tunneling crosses over to elastic co-tunneling.

The Majorana bound states are zero energy solutions of the Bogoliubov-de Gennes equations, and have self-conjugate operators
\begin{equation}
\gamma_j = \sum_\sigma\int \de \vt{r}\,\left[u_j(\vt{r},\sigma)\Psi(\vt{r},\sigma)+u_j^*(\vt{r},\sigma)\Psi^\dagger(\vt{r},\sigma)\right]\,,
\end{equation}
where $j=l,r$ denotes the two Majoranas at opposite ends of the wire, and $u_j(\vt{r},\sigma)$ is a bound state wave function centered around the location $\vt{r}_j$ of either point contact. Both wave functions decay exponentially away from $\vt{r}_j$. The length scale for the decay is set by the effective superconducting coherence length $\xi$, which for a ballistic nanowire is equal to $\xi=v/\Delta$, where $v$ is the renormalized Fermi velocity for states in the proximitized nanowire. The latter can be estimated from Eq.~\eqref{eq:level_spacing} as $v\approx Z_0\alpha$.

Provided that all relevant energy scales are smaller than the gap $\Delta$, one may replace the full tunneling Hamiltonian of Eq.~\eqref{eq:h_tunn} with a low-energy version which only takes into account tunneling from the lead into the nanowire via the Majoranas. In this approximation, the electron field operator is written as $\Psi(\vt{r},\sigma)=\sum_j u_j(\vt{r},\sigma)\,\gamma_j +\dots$ and the tunneling Hamiltonian of Eq.~\eqref{eq:h_tunn} becomes
\begin{equation}
H_\textrm{tunn} = \sum_{j,p\sigma}\,[W_j\,u^*_j(\vt{r}_j,\sigma)\,\phi_p(\vt{r}_j)\gamma_j\,c_p\,\hat{N}^+ +\textrm{H.c.}]+\dots\,,
\end{equation}
where $\hat{N}^+$ is a raising operator for the number of electron charges in the proximitized nanowire, which is included to make the tunneling Hamiltonian explicitly charge-conserving. The dots in the equation above indicate omission of states above the gap.

The localized nature of the Majorana bound states has important consequences for the magnitude of the level broadening $\Gamma_\textrm{Maj}$ of the zero-energy state induced by the presence of the contacts. Indeed, the normalization for the Majorana wave function requires $|u(\vt{r}_j,\sigma)|^2 \sim (Z_0/\xi)$, where the factor $Z_0$ again takes into account the reduced weight of wave functions in the nanowire due to the coupling to the superconductor. Hence, for a ballistic nanowire, we obtain the following estimate:
\begin{align}\nonumber
\Gamma_\textrm{Maj} &= \pi \sum_{j,p\sigma} W^2_j |u_j(\vt{r}_j,\sigma)|^2 |\phi_p(\vt{r}_j)|^2 \delta(\xi_p) \simeq \\
&\simeq \frac{(g_l+g_r)\,\Delta}{8\pi}\,.\label{eq:Gamma_Maj}
\end{align}
For a given sample, the value of $\Gamma_\textrm{Maj}$ may be affected by mesoscopic fluctuations, and in particular by the microscopic details of the portion of the nanowire close to the contacts. However, the crucial fact is that the relevant energy scale for the broadening is the gap $\Delta$, rather than the level spacing $\delta$.

\begin{figure}[!t]
\begin{center}
\includegraphics[width=\columnwidth]{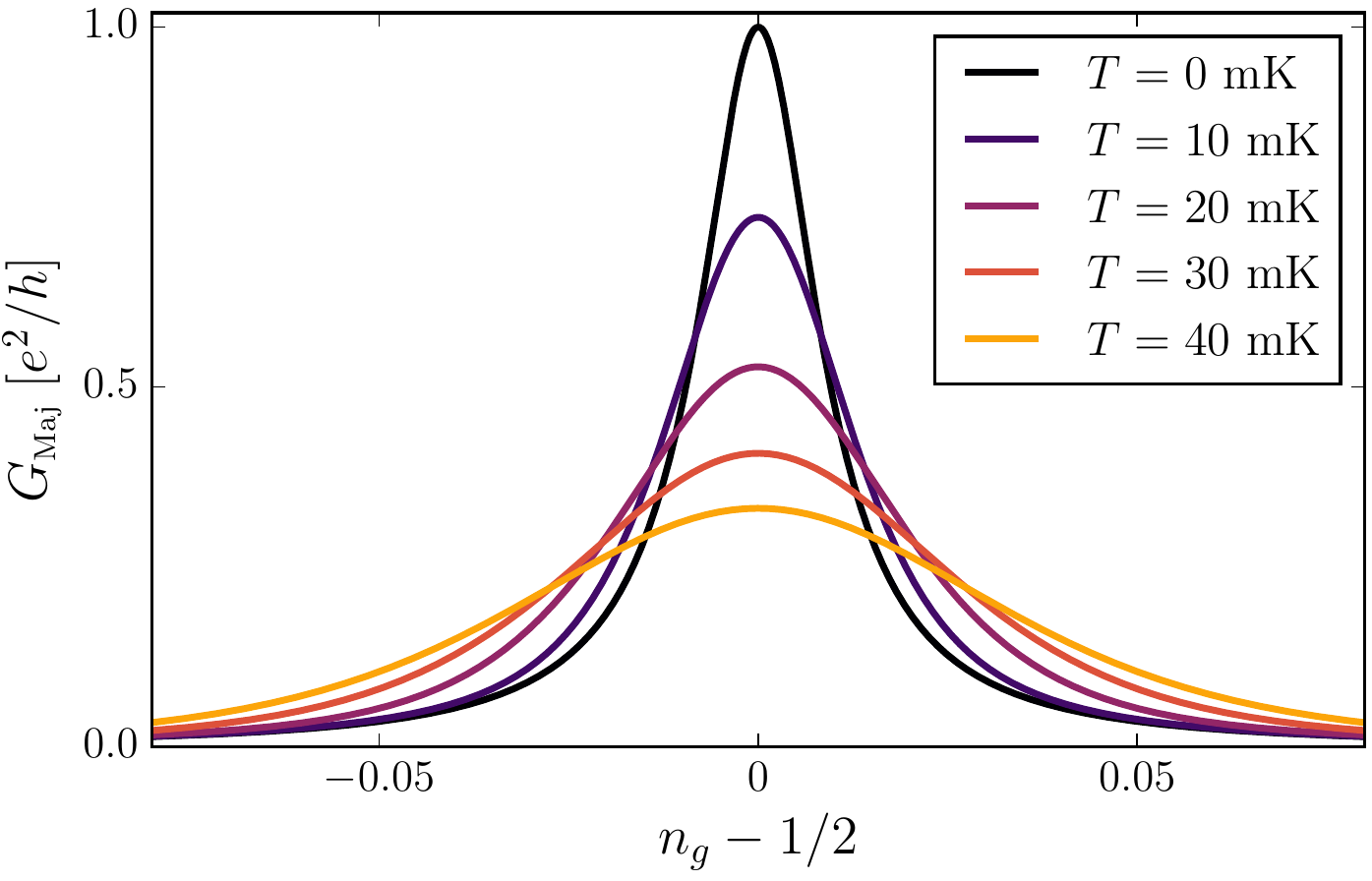}
\caption{Plot of the conductance $G_\textrm{Maj}$ due to resonant tunneling through the Majorana bound states at $B>B_c$, Eq.~\eqref{eq:g_maj}, at different temperatures $T$. We used the same parameter as in Fig.~\ref{fig:g_seq}.\label{fig:g_majorana}}
\end{center}
\end{figure}

The calculation of the conductance is equivalent to that of the resonant tunneling of electrons via a double barrier hosting a single bound state with energy $E_1-E_0 = E_c(2n_g-1)$. The probability for such a process is described by the Breit-Wigner formula,
\begin{equation}
\abs{A_p}^2 = \frac{g_l g_r}{4(2\pi)^4} \frac{\delta_l\delta_r\,\Delta^2}{(E_1\!-\!E_0\!-\!\xi_p)^2+(g_l\!+\!g_r)^2\Delta^2/(8\pi)^2}\,,
\end{equation}
with $\xi_p$ the energy of the initial state in the leads. The summation of the probability over the states in the leads yields the following integral expression for the linear conductance,
\begin{equation}\label{eq:g_maj_int}
G_\textrm{Maj} = \frac{2\pi e^2}{\hbar}\frac{1}{4T\,\delta_l\delta_r}\int_{-\infty}^\infty\frac{\de \xi_p\,\abs{A_p}^2}{\cosh^2(\xi/2T)}\,.
\end{equation}
In the limit $T\ll\Gamma_\textrm{Maj}$ one obtains
\begin{equation}\label{eq:g_maj_zero_T}
G_\textrm{Maj} = \frac{e^2}{h}\,\frac{g_lg_r}{4(2\pi)^2}\,\frac{\Delta^2}{4E_c^2(n_g\!-\!\tfrac{1}{2})^2 + (g_l\!+\!g_r)^2\Delta^2/(8\pi)^2}\,.
\end{equation}
This is a resonant peak centered at $n_g=\tfrac{1}{2}$, with height $(e^2/h)\, 4 g_lg_r/(g_l+g_r)^2$ and half-width at half-maximum $(g_l+g_r)\Delta/16\pi E_c$. Note that the conductance maximum is $e^2/h$ for a symmetric junction with $g_l=g_r$.

The integral in Eq.~\eqref{eq:g_maj_int} can be solved analytically also for a finite temperature,
\begin{align}\nonumber
& G_\textrm{Maj}(T) = \frac{e^2}{h}\,\frac{g_l g_r}{g_l+g_r}\frac{\Delta}{4\pi^2 T}\,\times\\\label{eq:g_maj}
&\quad \times\,\re\left[\psi'\left(\frac{1}{2}+\frac{(g_l+g_r)\Delta}{16\pi^2 T}-\frac{i(n_g-\tfrac{1}{2})E_c}{\pi T}\right)\right]\,,
\end{align}
where $\psi'(z)$ is the polygamma function of first order \cite{abramowitz}. The equation above describes the crossover from the zero temperature resonant peak to a temperature-broadened peak
\begin{equation}\label{eq:g_maj_finite_T}
G_\textrm{Maj} \simeq \frac{e^2}{h}\,\frac{g_l g_r}{g_l+g_r}\,\frac{\Delta}{8T}\,\frac{1}{\cosh^2[E_c(n_g-\tfrac{1}{2})/T]}
\end{equation}
at temperatures $T\gg\Gamma_\textrm{Maj}$. In Fig.~\ref{fig:g_majorana} we plot the conductance peak for several temperatures.

It is important to contrast the Coulomb blockade peak shapes in the case of tunneling via Majorana states with the peaks in the single-electron tunneling regime (see Sec.~\ref{sec:single_electron}). Unlike the latter, the conductance maxima we find here [see Eqs.~(\ref{eq:g_maj_zero_T}) and (\ref{eq:g_maj_int})] are symmetric with respect to the degeneracy point at any $T/\Gamma_\textrm{Maj}$. The difference stems from the different nature of the excitations spectra: at $B>B_c$ a substantial gap, $\sim\Delta (B)$, exists in each of the two states brought to degeneracy by adjusting the gate voltage $n_g$.

So far we have not considered exponentially small correction to the ground-state energy, which appears in a finite-length wire due to the hybridization of the two Majorana states~\cite{kitaev2001,cheng2009}. This correction will shift the conductance peak position while preserving the shape of the peak. The peaks symmetry substantiates the way small (smaller than the peak width) corrections to the peak positions were extracted in~\cite{albrecht2015}.

\section{Conclusions}

\begin{figure}[t!]
\begin{center}
\includegraphics[width=\columnwidth]{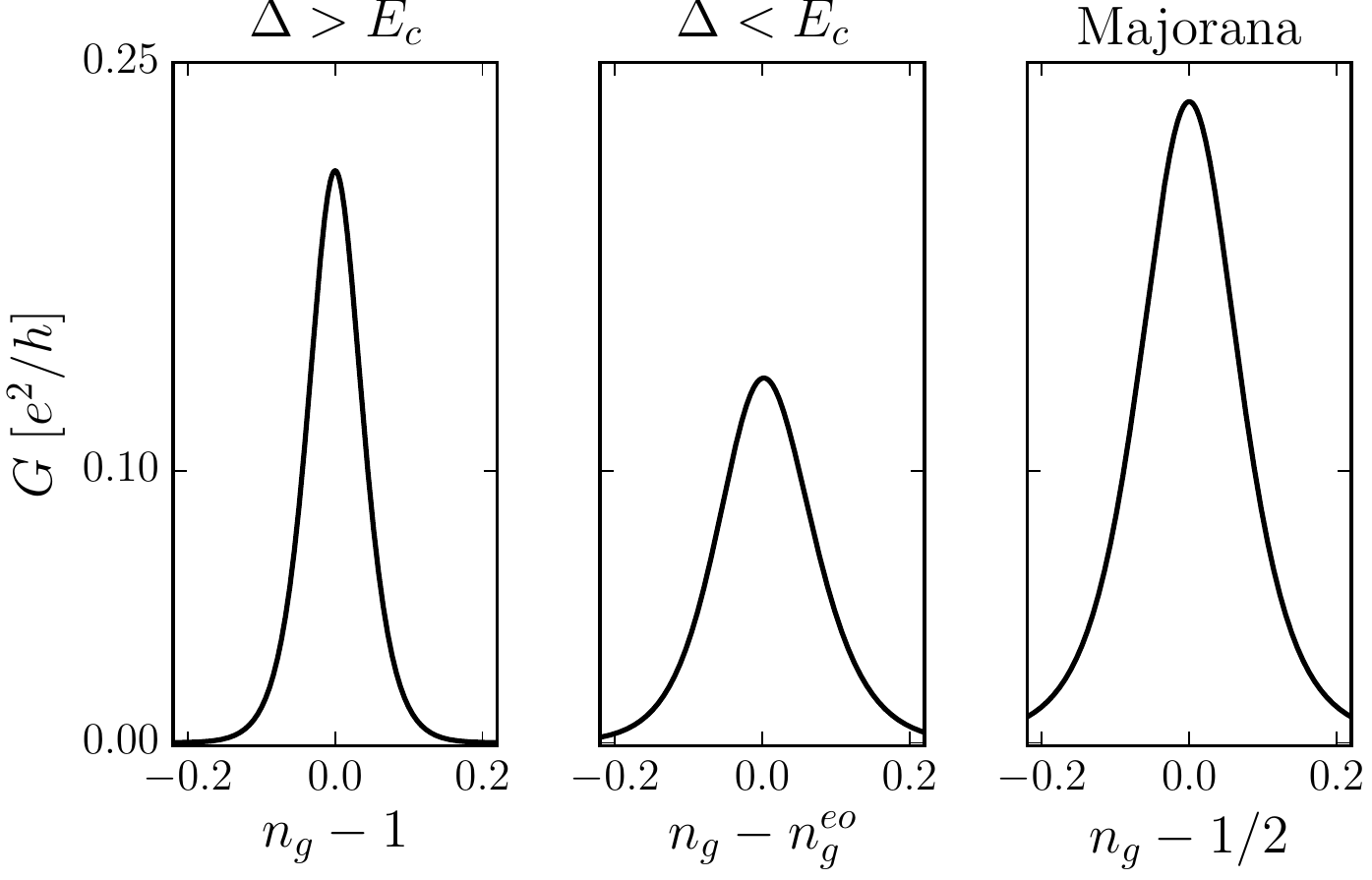}
\caption{Comparison between conductance peaks in the three different regimes of Coulomb blockade oscillations treated in this work: Andreev regime (left panel), single-electron tunneling regime (middle panel), and Majorana regime (right panel). We have used parameters comparable to those estimated for the device in Ref.~\cite{albrecht2015} which was closest to the weak tunneling regime considered in this paper: $\Delta_\textrm{Al}=180\,\mu$eV, $\Delta_0=130\,\mu$eV, $\Delta=30\,\mu$eV for the two rightmost panels, $E_c = 55\,\mu$eV, $T=50$ mK, $L=0.95\,\mu$m, $\alpha=8\,\mu$eV$\cdot\,\mu$m, $g_l=g_r=0.65$. The value of $g_l$ and $g_r$ was chosen to approximately match the height of the Andreev peak conductance observed for the $L=0.95\,\mu$m device in the experiment  \cite{albrecht2015} ($G_\textrm{2e}^\textrm{peak}\approx 0.2 \times e^2/h$).\label{fig:peak_comparison}}
\end{center}
\end{figure}

We have developed a quantitative theory of the two-terminal conductance through a proximitized nanowire in the Coulomb blockade regime. Inspired by the recent experiment~\cite{albrecht2015}, we have investigated the magnetic field dependence of the conductance and identified three distinct transport regimes which may occur upon increasing an external magnetic field $B$: Andreev transport regime (a), single-electron tunneling regime (b), and coherent transmission regime (c) through a Majorana zero-energy state which occurs when the system is driven into topological superconducting phase.

Using weak tunneling approximation, we have computed the shape of conductance peaks of Coulomb blockade oscillations for all three regimes, see Eqs.~\eqref{eq:G_2e}, \eqref{eq:sequential_peak}, and \eqref{eq:g_maj_finite_T}, and the corresponding Fig.~\ref{fig:peak_comparison}. Using our results, one can draw the following conclusions which are important for the interpretation of the experimental data~\cite{albrecht2015}.

First, the height of the conductance peaks is a non-monotonic function of magnetic field $B$ with the generic pattern of bright-dark-bright signals corresponding to (a), (b), and (c) regimes, respectively. In the limit of long wires, we predict that conductance should be suppressed in the single-electron tunneling regime (b) whereas Andreev (a) and Majorana (c) contributions to the conductance should remain finite.

Second, the width of the Coulomb peaks provides additional information regarding the nature of transport mechanisms for a given magnetic field. Upon lowering the temperature, the Coulomb blockade peaks widths in the regime (c) saturate, see Eq.~\eqref{eq:g_maj_zero_T}. At higher temperatures, the peak widths are proportional to temperature $T$, being limited by thermal activation in each of the three regimes; the width in the regime (c) is twice bigger than in the regime (a) where conduction is facilitated by hopping of electron pairs.

Third, the relative height of an Andreev peak should increase with increasing the conductances of the point contacts whereas the ratio of the Coulomb blockade peaks in the single-electron tunneling and activation-limited Majorana regimes is independent of  $g_l$ and $g_r$, cf. Eqs.~(\ref{eq:G_2e}), (\ref{eq:sequential_peak}), (\ref{eq:g_seq_peak}), and (\ref{eq:g_maj_finite_T}). For realistic physical parameters, we find that Andreev conductance should be smaller than the conductance in the topological regime, see Fig.~\ref{fig:sketch_peaks}, whereas the experimental findings~\cite{albrecht2015} are the opposite. This quantitative discrepancy might be due to our single-channel approximation. It is likely that the nanowire might have a few transverse channels, which would not affect the conductance in regime (c), while enhancing the regime (a) conductance.

Finally, we find that Coulomb blockade peaks in the Majorana regime (c) are described by an even function (a Lorentzian at low $T$) centered, at any temperature, exactly at the point of degeneracy of two ground states differing by single-electron charge. This should be contrasted with the conductance in the single-electron tunneling regime (b) where the peak positions are $T$-dependent and shifted away from the degeneracy points,  while the peaks shape is skewed with respect to their maxima. Thus, we find that in the Majorana regime (c) the position of the Coulomb blockade peaks, even if those are thermally-broadened, can be used as a sensitive probe of the ground-state degeneracy splitting due to a finite length of a nanowire. In this sense, our finding corroborates the conclusions of Ref.~\cite{albrecht2015}.

We note that in the experiment~\cite{albrecht2015}, the dimensionless conductances $g_l, g_r$ were set to quite large (i.e. order one) values; a systematic investigation of the two-terminal conductance as a function of the left/right tunnel barriers transmission coefficients would be very useful. On the theory side, it is desirable to extend the consideration to include the effect of almost-open junctions, higher channel number, and mesoscopic fluctuations.

\acknowledgments

We acknowledge stimulating discussions with P.W.~Brouwer, C.~M.~Marcus, K.~Flensberg, and A.~Kamenev.  BvH was supported by ONR Grant Q00704. LG acknowledges the support by DOE contract DEFG02-08ER46482. RL wishes to acknowledge the hospitality of the Aspen Center for Physics supported by NSF Grant \#1066293.

\begin{appendix}
\section{Level spacing of a proximitized nanowire}
\label{app:level_spacing}

In this appendix we derive Eq.~\eqref{eq:level_spacing} for the level spacing of a proximitized Rashba nanowire close to the topological phase transition at $B=B_c$. We begin from the known expression \cite{potter2011,stanescu2011} for the single-particle Green's function $G(k,E)$ for an electron propagating along the nanowire with momentum $k$ and energy $E$,
\begin{equation}\label{eq:GkE}
G(k,E) = \frac{Z(E)}{E- Z(E)\,H(k) + [1-Z(E)]\,\Delta_\textrm{Al}\,\tau_1}\,.
\end{equation}
Here, $H(k)$ is the Hamiltonian for the nanowire in the absence of the superconductor, $\Delta_\textrm{Al}$ is the superconducting gap in Al, $\tau_1$ is the first Pauli matrix in Nambu space, and $Z(E)$ is a renormalization factor due to the coupling with the superconductor:
\begin{equation}\label{eq:z}
Z(E) = \frac{1}{1+\Gamma/\sqrt{\Delta_\textrm{Al}^2-E^2}}\,.
\end{equation}
$\Gamma$ is an unknown parameter with the physical dimension of energy, which measures the coupling strength between the wire and the superconductor. $Z(E)$ can be interpreted as the fraction of time that a particle with energy $E<\Delta_\textrm{Al}$ spends in the semiconducting nanowire, as opposed to the superconductor.

For simplicity, we take for $H(k)$ the standard Hamiltonian of a single-channel Rashba wire in a magnetic field \cite{oreg2010,lutchyn2010}:
\begin{equation}
H(k) = [\zeta(k) + \alpha k \sigma_2]\,\tau_3 + g \mu_B B\, \sigma_3\,.
\end{equation}
Here, $\zeta(k) = k^2/2m-\mu$, $m$ is the effective mass in the semiconductor, $\alpha$ is the strength of the spin-orbit coupling, $g$ the g-factor in the nanowire, $\mu_B$ the Bohr magneton, $B$ is the magnetic field, and $\sigma$ and $\tau$ are Pauli matrices in spin and Nambu space respectively.

The energy spectrum of the proximitized nanowire can be found by solving the equation
\begin{equation}
\det\,[G^{-1}(k,E)] = 0\,.
\end{equation}
The determinant can be calculated explicitly. It gives the following equation for $E$ \cite{stanescu2011}:
\begin{align}\nonumber
\frac{E^2}{Z^2(E)} &= \frac{\Delta_\textrm{Al}^2[1-Z(E)]^2}{Z^2(E)} + [V_z^2 + k^2\alpha^2 + \zeta^2(k)]\\
&\qquad \pm 2\sqrt{\zeta^2(k)\,(V_z^2+k^2\alpha^2)+\frac{V_z^2\Delta_\textrm{Al}^2[1-Z(E)]^2}{Z^2(E)}}
\label{eq:spectrum_with_self_energy}
\end{align}
with $V_z = g \mu_B B$.

The level spacing $\delta$ of a nanowire of length $L$ will depend crucially on the strength of the proximity effect. We do not want to perform a systematic study of the level spacing as a function of all the parameters, but rather to obtain an estimate for $\delta$ without making any assumption on the value of $\Gamma$, which is an unknown parameter not easy to control in experiment nor to extract from experimental data. Our strategy is to first find an equation for $\Gamma$ in terms of the observable quantity $\Delta_\textrm{Al}$ and $\Delta_0$, the induced gap at $B=0$.

To obtain such equation, we focus on the lowest energy branch in Eq.~\eqref{eq:spectrum_with_self_energy}.  The relevant gap in the spectrum is expected to be at $k=0$, at least up to values of magnetic field larger than $B_c$, and to reach its optimal value at $\mu=0$. Hence we set $E=\Delta_0$, $k=0$, $\mu=0$ and $B=0$ in Eq.~\eqref{eq:spectrum_with_self_energy} and we find
\begin{equation}\label{eq:gamma_vs_induced_gap}
\Gamma =\Delta_0\,\sqrt{\frac{\Delta_\textrm{Al} +\Delta_0}{\Delta_\textrm{Al}-\Delta_0}}\,.
\end{equation}
This equation establishes the sought relation between $\Gamma$ and $\Delta_0$ and it is valid both for weak proximity (that is, $\Delta_0\ll \Delta_\textrm{Al}$ or equivalently $\Gamma\ll \Delta_\textrm{Al}$) and strong proximity (that is, $\Delta_0\to \Delta_\textrm{Al}$ or equivalently $\Gamma\gg \Delta_\textrm{Al}$).

It is worth stopping one moment to analyze Eq.~\eqref{eq:gamma_vs_induced_gap}. In both limits of weak and strong proximity we can find approximate expressions for $\Delta_0$ as a function of $\Gamma$ by expanding the right hand side around $\Delta_0 = 0$ and $\Delta_0=\Delta_\textrm{Al}$ respectively. For weak proximity one obtains $\Delta_0\approx \Gamma$, while for strong proximity $\Delta_0 \approx \Delta_\textrm{Al}(1 - 2\Delta_\textrm{Al}^2/\Gamma^2)$. As a remark, we want to stress the difference between this result and the expression $\Delta_0 = (\Gamma \Delta_\textrm{Al})/(\Gamma+\Delta_\textrm{Al})$ which is often used in the literature, and which can be obtained via the same derivation but by replacing $Z(E)$ with its value at $E=0$ in Eq.~\eqref{eq:spectrum_with_self_energy}. The latter expression gives the wrong asymptotic expansion for $\Gamma\gg \Delta$, and in fact the induced gap approached $\Delta_\textrm{Al}$ faster upon increasing $\Gamma$.

Let us now consider $B=B_c$, with $\Delta(B_c)=0$ and $\mu=0$. In this regime it is indeed appropriate to replace $Z(E)$ with its value at $E=0$, $Z_0=(1+\Gamma/\Delta)^{-1}$, in Eq.~\eqref{eq:spectrum_with_self_energy}. Using Eq.~\eqref{eq:gamma_vs_induced_gap} to replace $\Gamma$ in $Z_0$, we obtain the expression quoted in Eq.~\eqref{eq:Z0} of the main text. Neglecting quadratic terms in $k$ in $H(k)$, that is focusing on momenta $k\ll m\alpha$, Eq.~\eqref{eq:spectrum_with_self_energy} simply gives
\begin{equation}
E = Z_0\,\alpha k\,.
\end{equation}
For a wire of length $L$ the momentum is quantized in multiples of $\pi/L$, leading to $\delta = Z_0 \pi \alpha/L$, Eq.~\eqref{eq:level_spacing} of the main text.

\end{appendix}

\bibliography{draft}

\end{document}